\documentclass[12pt,a4paper]{article}
\usepackage[utf8]{inputenc}
\usepackage{amsmath}
\usepackage{placeins}
\usepackage{url}
\usepackage{svg}
\usepackage{natbib}
\usepackage{titlesec}
\usepackage{color,xcolor,setspace,geometry}
\setcitestyle{authoryear,open={(},close={)}}
\usepackage{cite}
\usepackage{soul}
\usepackage{mathtools}
\usepackage{bbm}
\usepackage{dsfont}
\soulregister\citep7
\soulregister\cite7
\soulregister\citet7
\soulregister\citealp7
\soulregister\ref7
\usepackage{amsfonts}
\usepackage{tikz}
\usetikzlibrary{shapes}
\usetikzlibrary{plotmarks}
\usetikzlibrary{tikzmark,matrix,calc}
\usetikzlibrary{arrows.meta,calc,decorations.markings,math,arrows.meta}
\usepackage{amssymb}
\usepackage{multirow}
\usepackage{hyperref}
\usepackage{todonotes}
\usepackage{graphicx}
\usepackage{array}
\usepackage{bm}
\usepackage{arydshln}
\usepackage[final]{changes}
\titleformat*{\section}{\large\bfseries}
\titleformat*{\subsection}{\bfseries}

\providecommand{\keywords}[1]
{
  \small	
  \textbf{\textit{Keywords---}} #1
}

\usepackage{natbib}
\usepackage{datatool}
\usepackage{etoolbox}

\DTLnewdb{bibnotes}

\makeatletter
\patchcmd{\@lbibitem}
  {\item[\hfil\NAT@anchor{#2}{\NAT@num}]}
  {
    \item[\hfil\NAT@anchor{#2}{\NAT@num}]
    \DTLforeach[\DTLiseq{\mylabel}{#2}]{bibnotes}{\mylabel=mylabel,\mynote=mynote}{\text{\mynote}}
  }{}{\message{^^JPatching failed^^J}}

\newcommand{\given}{\mid}

\DeclareMathOperator*{\argmax}{arg\,max}

\makeatother
\begin{document}
\begin{spacing}{1.35}

\listofchanges  
%\title{Capturing dependencies nonparametrically in multivariate hidden Markov models}
\title{Integrating Unsupervised and Supervised Learning for the Prediction of Defensive Schemes in American football}
\author{Rouven Michels\thanks{corresponding author: rouven.michels@tu-dortmund.de} \thanks{TU Dortmund University, Germany}, Robert Bajons\thanks{WU Vienna, Austria}, Jan-Ole Fischer\thanks{Bielefeld University, Germany}}
\date{}

\maketitle

\begin{abstract} 
\noindent
Anticipating defensive coverage schemes is a crucial yet challenging task for offenses in American football. Because defenders’ assignments are intentionally disguised before the snap, they remain difficult to recognize in real time. To address this challenge, we develop a statistical framework that integrates supervised and unsupervised learning using player tracking data. Our goal is to forecast the defensive coverage scheme --- man or zone --- through elastic net logistic regression and gradient-boosted decision trees with incrementally derived features. We first use features from the pre-motion situation, then incorporate players’ trajectories during motion in a naive way, and finally include features derived from a hidden Markov model (HMM). Based on player movements, the non-homogeneous HMM infers latent defensive assignments between offensive and defensive players during motion and transforms decoded state sequences into informative features for the supervised models. These HMM-based features enhance predictive performance and are significantly associated with coverage outcomes. Moreover, estimated random effects offer interpretable insights into how different defenses and positions adjust their coverage responsibilities.
\end{abstract}
\keywords{American football, forecasting, hidden Markov model, random effects, XGBoost}

\section{Introduction}

In American football, one of the most critical strategic questions an offense faces is identifying the defensive coverage scheme before the snap. Whether the defense is aligned in man- or zone coverage dictates the effectiveness of route combinations, blocking assignments, and ultimately the success of an offensive play. A central tool offenses employ to gain information about the defensive scheme is pre-snap motion. By moving receivers, tight ends, or running backs, coordinators seek to trigger visible reactions from defenders that provide cues about their coverage responsibilities. If a cornerback follows a motioning receiver across the entire field, this may indicate a man-to-man coverage scheme; if defenders pass on offensive players and shift in the direction of the motion, zone coverage becomes more likely. Thus, pre-snap motion serves as a diagnostic tool to gain information on the defensive scheme.

Historically, the interpretation of pre-snap movements has been almost entirely visual. Coaches and analysts rely on film study to infer how defenses react to offensive formations and motions. While this process has proven effective in practice, it remains subjective.
The increasing availability of player tracking data offers a way to overcome these limitations. These data include spatial information on player movements, thereby fundamentally changing how defensive coverages --- and football situations more broadly --- can be evaluated. 
Over the past decade, the integration of play-by-play and tracking data has reshaped multiple domains of football analytics, for example, fourth-down decision-making \citep{sandholtz2024learning, brill2025analytics}, the in-game valuation of plays \citep{yurko}, the understanding of pass rushes \citep{nguyen}, and the contribution of tackles \citep{bajons2025pep}. A notable step towards the prediction of coverage schemes was made by \citet{dutta2020unsupervised}, who employed mixture models to infer defensive assignments from player tracking data. While this approach demonstrated the potential of data-driven inference for coverage identification, it did not incorporate information on pre-snap motion. However, this phase may hold valuable latent patterns that can be statistically inferred and leveraged for improved coverage identification.

To fill this gap, we aim to predict the overall coverage scheme through a hybrid statistical framework combining unsupervised and supervised learning using play-by-play and tracking data from the first nine weeks of the 2024 NFL season. Specifically, we classify man- or zone coverage with an elastic net logistic regression and gradient-boosted decision trees based on contextual and spatial features from pre- and post-motion phases. 
To identify coverage patterns based on observed player movements while accounting for persistence in latent guarding assignments, we employ hidden Markov models (HMMs).
HMMs are a versatile statistical framework for modeling time series driven by underlying, unobserved states and have been applied to various fields, including finance \citep{liu2012stock, zhang2019high}, ecology \citep{mcclintock2020uncovering, mews2022multistate}, and medicine \citep{amoros2019continuous, soper2020hidden}.
Moreover, latent-state time-series models have been employed across multiple sports analytics domains to uncover unobserved structures, such as tactics (\citealp{otting2023football, adam2024markov}) or hot hand and momentum effects (\citealp{albert93, michels2023bettors, winkelmann2025momentum}).

In our case, we employ a non-homogeneous HMM to infer latent guarding assignments between defensive and offensive players during the pre-snap motion phase. For this, we model the defenders’ trajectories in each play conditional on the offensive players' positions. Local decoding of each defender’s time series provides probabilistic information on the guarding assignments from each defensive player to each offensive player in each play. 
The results from the HMM are then transferred into suitable and interpretable summary statistics and incorporated as features in the supervised learning models.
We demonstrate its usefulness in two complementary ways. First, we show that adding HMM-derived features enhances the prediction of man- or zone coverage with regard to several common loss metrics compared to models solely relying on naive pre- and post-motion features. %By devising a suitable cross-fitting procedure, we are able to reveal that the models incorporating HMM features improve out-of-sample performance compared to models solely relying on naive pre- and post-motion features. 
Second, by leveraging recently established non-parametric conditional independence tests and more specifically the Generalized Covariance Measure (GCM) test \citep{Shah20gcm}, we show that the HMM features are significantly associated with the coverage outcome. Furthermore, using results from \citet{bajons2025rGAX}, the GCM test provides interpretable insights into the influence of the HMM features on the outcome. These results provide further evidence of the usefulness of our approach. Finally, we use the best-performing coverage model to examine team behavior and motion patterns. More specifically, we analyze which teams are able to exploit motion in order to more effectively determine defensive coverage schemes. 

The paper is organized as follows. Section~\ref{sec:data} describes the data. Section~\ref{sec:methods} presents the methodology, including the unsupervised and supervised modeling components. Section~\ref{sec:results} reports the results. Section~\ref{sec:discussion} concludes with a discussion.

\section{Data}
\label{sec:data}
\begin{figure}[t!]
    \centering
    \includegraphics[width=\linewidth]{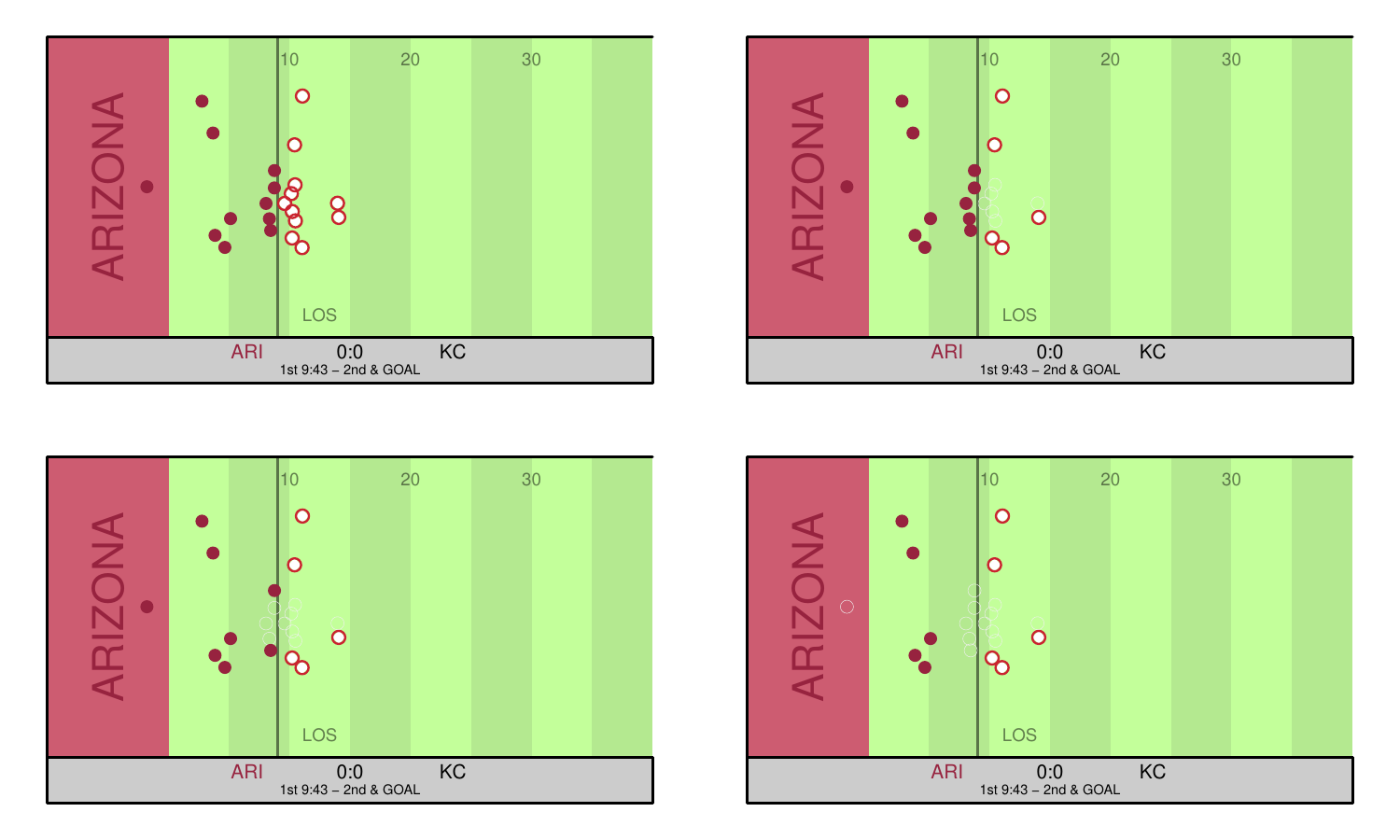}
    \caption{This figure illustrates the step-by-step pre-processing for a game between the Arizona Cardinals and the Kansas City Chiefs. From top left to bottom right, we first display all players, then exclude offensive non-skill players, next omit defensive linemen, and finally remove pass-rushing linebackers and deep safeties. Grey circles mark the players excluded during each pre-processing step.}
    \label{fig:prepro}
\end{figure}

The analysis in this paper is based on data provided as part of the NFL Big Data Bowl (BDB) 2025, specifically tracking data, sampled at 10 Hz, and play-by-play data from the first nine weeks of the 2023 NFL season. While the former provides information on the locations of every player on the field during a play, the latter contains general, contextual information on the respective plays. As we are interested in analyzing pre-snap player movements, we concentrate on plays that contain any pre-snap motion from the offense. Moreover, we omit plays with two quarterbacks and bunch formations. Ultimately, we end up with $M = 3963$ offensive plays in total, of which the different defenses played $2980$ in zone and $983$ in man coverage. 

Our modeling framework is structured into three stages: a pre-motion model, a naive post-motion model, and a post-motion model including HMM features. Thereby, the amount of available information progressively increases. Each subsequent stage builds upon the data used in the previous models.
Starting with the creation of the pre-motion data set, we first extract information from the play-by-play data, specifically the current quarter, down, yards to go, yardline, home and away scores, the remaining seconds in the current half, and, most importantly, whether the defense played man- or zone coverage\footnote{Note that this information is itself provided by the company Pro Football Focus as part of the NFL BDB data.}. In addition, we create different features derived from the tracking data that describe the situation before any player motions. In particular, we first compute the area spanned by the convex hull of all players from both teams, as well as the largest y-distance (i.e., the width of the hull) and the largest x-distance (i.e., the length of the hull) of both convex hulls. 
Subsequently, we concentrate on the offensive players that can potentially move before the snap, i.e., exclude offensive linemen and the quarterback. This yields five offensive skill players (an arbitrary combination of running backs, full backs, tight ends, and wide receivers) for each individual play. 
To enable direct man-to-man assignments later in the HMM framework, we match the number of defensive players with the five offensive players. For that, we omit defensive linemen (nose tackles, defensive tackles, and defensive ends) and those players that are tagged as outside linebackers but align directly at the line of scrimmage as pass rushers (players such as Micah Parsons).
Afterwards --- if more than five defenders remain --- we select the five defenders closest to those offensive players, typically excluding deep safeties, which leads to $5 \times M = 19,815$ time series. Figure~\ref{fig:prepro} exemplifies all these pre-processing steps for a play between the Arizona Cardinals and the Kansas City Chiefs in Week 1, which will serve as a running example throughout this manuscript.
From these remaining ten players, we derive twenty features related to their (standardized) position. A detailed description is provided in Section~3 of the  Supplementary Material.

%Each observation corresponds to a unique frame within a play and includes the (x,y) coordinates of every player on the field, their speed, acceleration, orientation, and team affiliation, as well as contextual information such as down, distance, and game situation.
In a second step, to create the naive post-motion data set, we capture post-motion information purely from tracking data. In particular, we introduce six features: for both the offense and defense, we compute the maximum distance along the x- and y-axes, as well as the total
distance traveled by all players between the start of the motion and the snap. These features are added to the those used in the pre-motion model.

Finally, the HMM data set combines all previously derived features with additional variables obtained from suitable summary statistics of decoded state sequences. Information on how this dataset was created is provided in Section~\ref{sec:summary}. Throughout this process, no modifications were made to the raw time-series structure of the tracking data so as to preserve its underlying temporal structure (\citealp{michels2024combination}).

\section{Methods}\label{sec:methods}

Our modeling framework is designed in two layers. On the outer level, we employ two supervised learning models to predict the defensive coverage scheme based on contextual and spatial features described in the previous section. However, one of the most informative indicators for the defensive scheme, the assignments of defenders to offensive players, is not directly observable. Thus, we use a hidden Markov model as the inner component of our statistical framework to generate additional latent-state features that serve as inputs for the supervised models.

\subsection{Unsupervised learning}
As the unsupervised learning part, we use a hidden Markov model to infer guarding assignments between offensive and defensive players. HMMs are particularly well-suited for this task since these guarding assignments are not directly observable but must be inferred from the defenders' responses to the offensive players' movements. 
In this section, we describe our formulation and estimation of the HMM and explain how decoded state probabilities are incorporated as features into the supervised models.

\subsubsection{HMM formulation}

An HMM consists of two stochastic processes: an observed process $\{Y_t\}_{t=1}^{T}$ and an unobserved first-order Markov chain $\{ S_t\}_{t=1}^{T}$, with $S_t \in \{1,\ldots,N\}$ where $N$ is the number of states. The connection of both processes arises from the assumption that the observations $y_t$ --- in our case, literally the (vertical) $\mathrm{y}$-coordinates of the individual defenders in each play --- are generated by a state-dependent distribution that, in a basic model, only depends on the currently active state $S_t$. More formally, for each defender's trajectory $\{Y_t\}$, we assume that
\begin{equation}
\label{ind}
f(y_t \mid s_1, \ldots, s_t, y_1, \ldots, y_{t-1},y_{t+1},\ldots,y_{T_m}) = f(y_t \mid S_t = s_t),
\end{equation}
where $s_t \in \{1, \ldots, N\}$ and $t = 1, \ldots, T_m$, with $m = 1, \ldots, M$ specifying the m-th play. Note that, according to this model formulation, we model the guarding assignments of every individual defender. Thus, we assume that the trajectories of individual defenders are statistically independent from those of their teammates. Hence, it is possible that multiple defenders are assigned to one of the $N$ offensive players in a specific play $m$ (which, in reality, does happen), but not vice versa. 
We set $N=5$ due to the predetermined number of five offensive skill players that can potentially be guarded. 

\begin{figure}[t!]
    \centering
    \includegraphics[width=\linewidth]{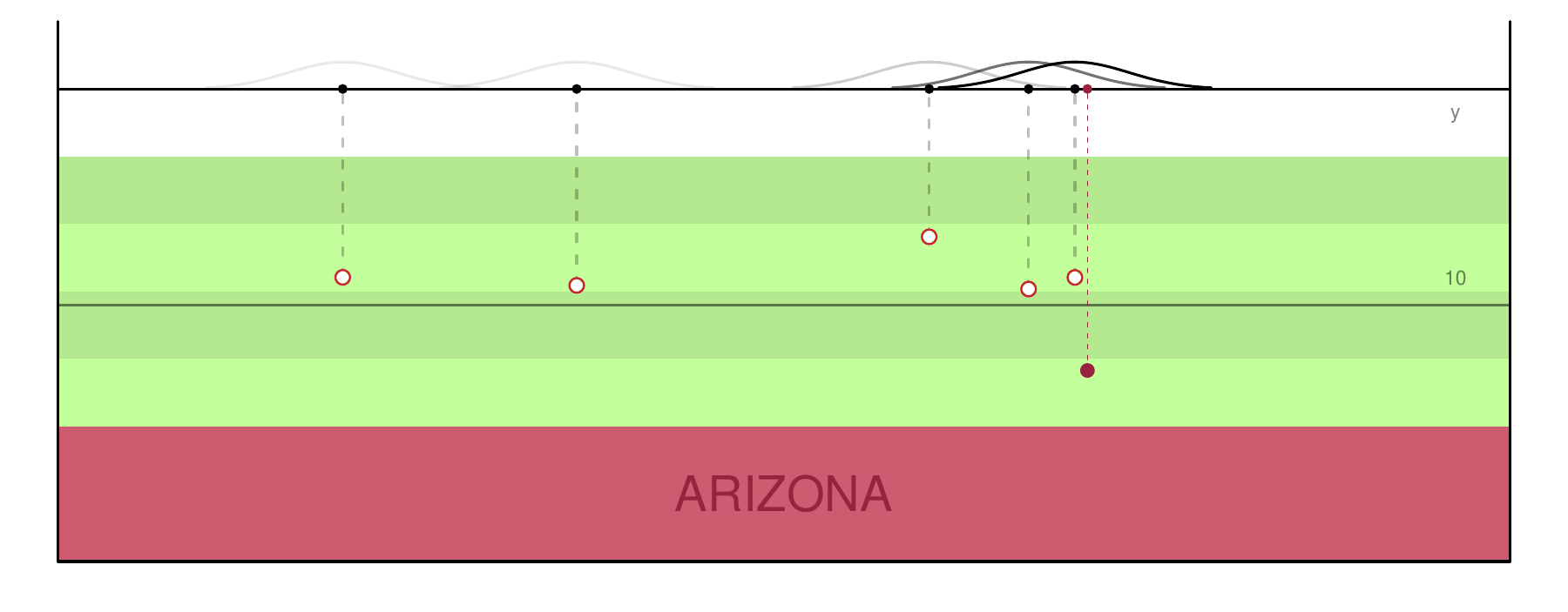}
    \caption{This figure illustrates the HMM modeling approach. Specifically, we project the $(x,y)$-coordinates of the players onto a single $\mathrm{y}$-axis. We assume that the observations $y_t$, i.e., the defensive players' $\mathrm{y}$-coordinate, is generated by one of the five Gaussian state-dependent distributions, whose mean equals the $\mathrm{y}$-coordinate of the five offensive players.}
    \label{fig:HMMexpl}
\end{figure}

We operationalize Eq.~\ref{ind}, by assuming that, at each time point $t$, every defender's $\mathrm{y}$-coordinate is a realization from a mixture of Gaussian distributions, where each mixture component represents the location of one offensive player (see \citealp{franks2015} for a similar approach in basketball). 
More formally, a natural candidate for the state-dependent distributions is
\begin{equation}
\label{eqn:state_dep}
  f(y_t \mid S_t = j) \sim \mathcal{N}(\mu_{t,j}, \sigma^2), \quad j = 1, \ldots, 5,  
\end{equation}
where the standard deviation $\sigma$ is estimated and the mean
% $$\mu_{t,j} = y_{\text{off}_{j, t}},$$ 
$$\mu_{t,j} = \mathrm{y}_t(\text{off}_j),$$ 
which denotes the $\mathrm{y}$-coordinate of offensive player $j$ at time $t$. The idea behind this modeling framework is that we expect defenders to mirror the offensive players at the line of scrimmage once teams come out of the huddle (see Figure~\ref{fig:HMMexpl}) and, in the case of tight coverage, also throughout the motion phase. 
A distinctive characteristic of the motion phase is that when offensive players move, defenders react to that movement. Thus, a more realistic model accounts for the time it takes for the defender to react to the offensive players' movement (see \citealp{wu2023evaluation} for a similar idea in soccer). Hence, we extend the basic modeling framework and allow for lagged observations within the mean of the state-dependent distribution, i.e., 
\begin{equation}\label{eq:lag}    
%\mu_{t,j} = \mathrm{y_\text{off}}^{j}_{t-l},
\mu_{t,j} = \mathrm{y}_{t-l}(\text{off}_j),
\end{equation}
with the lag $l$ being a proxy for the average reaction time of defenders. 
Given the 10 Hz temporal resolution of the data, $l$ can be interpreted as the number of tenths of a second it takes a defender to respond to an offensive player’s movement. For instance, $l = 5$ corresponds to an average reaction time of half a second.

\subsubsection{Modeling the underlying Markov chain}
\label{sec:MC_dets}

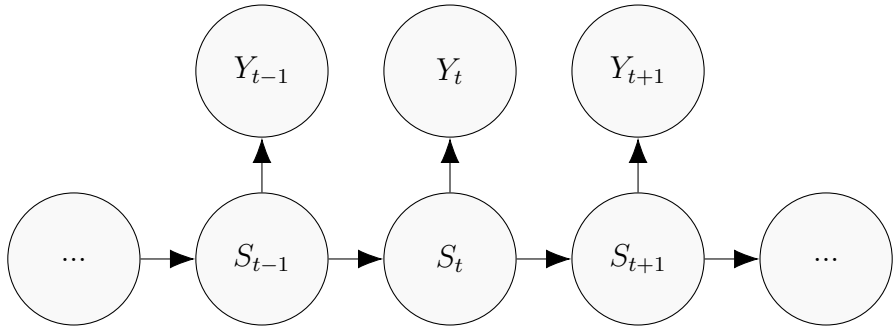
\begin{figure}[!b]
    \centering
	\begin{tikzpicture}
	\node[circle,draw=black, fill=gray!5, inner sep=0pt, minimum size=50pt] (A) at (2, -5) {$S_{t-1}$};
	\node[circle,draw=black, fill=gray!5, inner sep=0pt, minimum size=50pt] (A1) at (-0.5, -5) {...};
	\node[circle,draw=black, fill=gray!5, inner sep=0pt, minimum size=50pt] (B) at (4.5, -5) {$S_{t}$};
	\node[circle,draw=black, fill=gray!5, inner sep=0pt, minimum size=50pt] (C) at (7, -5) {$S_{t+1}$};
	\node[circle,draw=black, fill=gray!5, inner sep=0pt, minimum size=50pt] (C1) at (9.5, -5) {...};
	\node[circle,draw=black, fill=gray!5, inner sep=0pt, minimum size=50pt] (Y1) at (2, -2.5) {$Y_{t-1}$};
	\node[circle,draw=black, fill=gray!5, inner sep=0pt, minimum size=50pt] (Y2) at (4.5, -2.5) {$Y_{t}$};
	\node[circle,draw=black, fill=gray!5, inner sep=0pt, minimum size=50pt] (Y3) at (7, -2.5) {$Y_{t+1}$};
	\draw[-{Latex[scale=2]}] (A)--(B);
	\draw[-{Latex[scale=2]}] (B)--(C);
	\draw[-{Latex[scale=2]}] (A1)--(A);
	\draw[-{Latex[scale=2]}] (C)--(C1);
	\draw[-{Latex[scale=2]}] (A)--(Y1);
	\draw[-{Latex[scale=2]}] (B)--(Y2);
	\draw[-{Latex[scale=2]}] (C)--(Y3);
	\end{tikzpicture}
\caption{Dependence structure of the HMM used to model the defenders’ vertical positions $Y_t$, driven by the latent state variable $S_t$ that represents the currently guarded offensive player.}
\label{fig:HMM}
\end{figure}

Within our HMM (see Figure~\ref{fig:HMM} for a visualization of the basic structure), the unobserved state process is modeled as a first-order Markov chain. This is described by an initial distribution 
$$\boldsymbol{\delta}^{(1)} = \bigl( \Pr(S_1=1), \ldots, \Pr(S_1=5) \bigr),$$ 
that specifies the probability of every defender to guard the respective offensive players at the beginning of the time series, and a transition probability matrix (t.p.m.) 
$$\boldsymbol{\Gamma} = (\gamma_{ij}),\ \text{with} \  \gamma_{ij} = \Pr(S_t = j \mid S_{t-1} = i), \quad i,j = 1, \ldots, 5$$
The latter specifies the probability of a defender switching from guarding offensive player~$i$ at time~$t-1$ to offensive player~$j$ at time~$t$.
In a simple model, we would estimate a single, homogeneous transition probability matrix (t.p.m.). However, we seek to obtain a more realistic representation of the switching probabilities by allowing them to vary with covariates. 
Clearly, if a defender is currently guarding offensive player~$i$, a switch to guarding another offensive player~$j$ is more likely if those two are close in distance with respect to the $\mathrm{y}$-coordinate. Hence, we incorporate the pairwise distance between offensive players, i.e., $\vert \mathrm{y}_{t-l}(\text{off}_i) - \mathrm{y}_{t-l}(\text{off}_j) \vert$ for $i,j = 1, \ldots, 5$, as a covariate.
Further, we account for systematic heterogeneity across different positions, teams, and plays by including specific random effects in the state-transition probabilities (\citealp{maruotti2009semiparametric}). Thus, for the positional role $r \in \{\text{Cornerback, Inside/Outside/Middle Linebacker, Strong/Free Safety}\}$, the defensive team $d$ and the respective play $p$, this leads to mixed-effects linear predictors of the form
\begin{equation}\label{eq:re}    
\eta_{ij}^{(t,r,d,p)} = \beta_0 + \beta_1 \vert \mathrm{y}_{t-l}(\text{off}_i) -  \mathrm{y}_{t-l}(\text{off}_j) \vert + u_{r} + v_{d} + w_p,
\end{equation}
for $i \neq j$. Setting $\eta_{ii}^{(t,r,d,p)} \equiv 0$ for identifiability, we then express the transition probabilities of the unobserved Markov chain as functions of the above predictors using the inverse multinomial logistic link function
$$
\gamma_{ij}^{(t,r,d,p)} = \frac{\exp(\eta_{ij}^{(t,r,d,p)})}{\sum_{k=1}^N \exp(\eta_{ik}^{(t,r,d,p)})}, \quad i, j = 1, \dotsc, 5.
$$

For the random coefficients, we assume
$u_r \sim \mathcal{N}(0, \sigma_{\text{role}}^2)$ for all $r$ which captures deviations due to the specific defensive role $r$, $v_d \sim \mathcal{N}(0, \sigma_{\text{defense}}^2)$ for all $d$ which accounts for the defensive team-level tendencies and $w_p \sim \mathcal{N}(0, \sigma_{\text{play}}^2)$ for all $p$ which accounts for the play-level tendencies.
This hierarchical specification allows for partial pooling of information across positions, teams and plays, which is particularly beneficial given the limited number of observed plays. Consequently, the model remains flexible enough to learn meaningful structural differences between positions, teams and plays, while controlling for overfitting.

% In a second step, we account for systematic heterogeneity across different defensive positions and teams. Thus, we incorporate random effects into the transition probabilities and model the transition probabilities $\gamma_{ij}$ using a mixed-effects formulation:  

% \[
% \text{logit}(\gamma_{ij}^{(p,t)}) 
% = \beta_{ij} 
% + u_{ij}^{(\text{position}[p])} 
% + v_{ij}^{(\text{team}[t])},
% \]
% where $\beta_{ij}$ represents the global fixed effect, 
% $u_{ij}^{(\text{position}[p])} \sim \mathcal{N}(0, \sigma^2_{\text{position}})$ 
% captures deviations due to the position of player $p$, 
% and $v_{ij}^{(\text{team}[t])} \sim \mathcal{N}(0, \sigma^2_{\text{team}})$ 
% accounts for team-level tendencies.  

Beyond transition dynamics, the underlying Markov chain requires an initial state distribution, denoted by $\boldsymbol{\delta}^{(1)}$, which specifies the probability of each defender guarding a particular offensive player at the beginning of time series. In many HMM applications, this initial distribution is of limited practical relevance and is often replaced by the stationary distribution of the transition probability matrix.
However, in our setting this substitution would be inappropriate, as the initial distribution in the basic, homogeneous HMM is likely to differ substantially from the stationary one. In addition, an analytic solution for the stationary distribution of a non-homogeneous HMM can only be obtained for those that include periodic components in the underlying Markov chain (\citealp{koslik2025inference}), while a hypothetical stationary distribution would differ from the real one. This is particularly severe in our case, as the initial assignments carry meaningful information since they directly influence derived summary statistics such as the number of state transitions. Consequently, misspecifying $\boldsymbol{\delta}^{(1)}$ would distort these metrics and thereby have non-negligible effects on subsequent analyses. 
A potential alternative would be to estimate the initial distribution parameters in a data-driven manner. Yet we face $5 \times M = 5 \times 3963 =  19{,}815$ distinct time series, making individual estimation infeasible. To address this issue, we determine the initial state probabilities following a best-practice approach. Specifically, we model them as a function of observed spatial relationships between defensive and offensive players: higher initial state probabilities are assigned to offensive players that are locally close to the respective defenders, serving as a proxy for initial guarding assignments. Details on the specific implementation are provided in Section~1 of the Supplementary Material.

\subsubsection{Model fitting}

We fit our hidden Markov model under a frequentist framework by directly maximizing its marginal likelihood with random effects integrated out. Specifically, letting $f$ denote a general density, $\bm{y}$ the collection of all observations, and writing model parameters as subscripts, we have the marginal likelihood function
\begin{equation}
\label{eqn:marginal_lik}
    \mathcal{L}(\bm{\theta}) = f_{\bm{\theta}}(\bm{y}) = \iiint f_{\bm{\beta}, \sigma}(\bm{y} \vert \bm{u}, \bm{v}, \bm{w}) f_{\sigma_u}(\bm{u}) f_{\sigma_{v}}(\bm{v}) f_{\sigma_{w}}(\bm{w}) \; d\bm{u} \; d\bm{v} \; d\bm{w},
\end{equation}
where $\bm{\theta} = (\beta_0, \beta_1, \sigma, \sigma_u, \sigma_v, \sigma_w)$, and $\bm{u}$, $\bm{v}$, and $\bm{w}$ are vectors of random coefficients. 
$f_{\bm{\beta}, \sigma}(\bm{y} \vert \bm{u}, \bm{v}, \bm{w})$ is the density of the data given a particular value of the random effects, i.e.\ treating them as parameters, and $f_{\sigma_u}(\bm{u})$, $f_{\sigma_{v}}(\bm{v})$, and $f_{\sigma_{w}}(\bm{w})$ are mean-zero multivariate Gaussian densities with covariance matrices $\sigma_u^2 \bm{I}$, $\sigma_v^2 \bm{I}$ and $\sigma_w^2 \bm{I}$, respectively.
Due to the structure of the conditional density of $\bm{y}$ and the dimension of $\bm{u}$, $\bm{v}$ and $\bm{w}$, the above integral is analytically intractable, and hence needs to be evaluated approximately. This can be done using the so-called \textit{Laplace approximation}. A detailed explanation is given by \citet{kristensen2016tmb}, but the core idea is to approximate the log-integrand in \eqref{eqn:marginal_lik} (for fixed $\bm{\theta}$) around its mode (in $\bm{u}$, $\bm{v}$ and $\bm{w}$) by a quadratic function. Integrating the exponential of this quadratic function results in a Gaussian integral with a closed-form solution. Constructing the approximation around the mode means that for each evaluation of the marginal likelihood, an inner optimization over $\bm{u}$, $\bm{v}$ and $\bm{w}$ has to be performed.

In practice, we only need to specify the (negative) log-integrand in \eqref{eqn:marginal_lik}, that is, the \emph{negative joint log likelihood} of the data and the random effects as an \texttt{R} function. This is just the sum of the log likelihood of the data given the random coefficients and the logarithm of the Gaussian densities of the random coefficients. Based on this function, the \texttt{R} package \texttt{RTMB} \citep{rtmb} then generates a high-performance objective function approximating the logarithm of \eqref{eqn:marginal_lik} as well as an accompanying gradient using automatic differentiation. Internally, all computations are performed in \texttt{C++}. The returned objective function can then easily be numerically maximized using standard quasi-Newton numerical optimization routines such as the one implemented in \texttt{R}'s \texttt{nlminb} \citep{gay1990usage}, by supplying the objective function and its gradient.

%\subsubsubsectionline{Computing the joint log likelihood} 
Because we treat include a play-specific random effect, we treat the time series of different defenders within the same play as \emph{conditionally independent} and across plays as independent. Thus, the term $\log f_{\bm{\beta}, \sigma}(\bm{y} \vert \bm{u}, \bm{v}, \bm{w})$ %--- required as an \texttt{R} function for the approach outlined above --- 
is composed of a sum of individual HMM log likelihoods for each defender's movement. % within each play. 
In order to compute such a single term, we need to marginalize over all possible state sequences that could have generated the observations, which itself is a non-trivial task. However, the summation over latent states can be efficiently computed via the well- established \textit{forward algorithm} \citep{zucchini, mews2025build}, rendering the computational complexity linear in the number of observations. The idea of this recursive scheme is that we can traverse along the time series, at each point retaining information of the likelihood up to that point as well as probabilistic information on the current state given previous observations, which is made possible by the Markov assumption \citep{lystig2002exact}.

For a defender %$i$ 
of role $r$, team $d$ in play $p$, to save ink, we write the transition probability matrix $\bm{\Gamma}^{(t,r,d,p)} = (\gamma_{ij}^{(t,r,d,p)})$ simply as $\bm{\Gamma}^{(t)}$. 
Then, the conditional likelihood (given the position-, team- and play-specific random effects) of the corresponding defender's movement can be written down in closed form as
\begin{equation}
    \label{eqn:cond_lik}
    %\mathcal{L}(\bm{\theta} \vert \bm{u}, \bm{v}, \bm{w}) =
    \bm{\delta}^{(1)} \bm{P}(y_1) \bm{\Gamma}^{(2)} \bm{P}(y_2) \bm{\Gamma}^{(3)} \dotsc \bm{\Gamma}^{(T_m)} \bm{P}(y_{T_m}),
\end{equation}
where $\bm{\delta}^{(1)}$ is the play- and defender-specific initial distribution calculated as outlined in the previous section, and $\bm{P}(y_t) = \text{diag}\bigl(f(y_t \vert S_t = 1), \dotsc, f(y_t \vert S_t = 5)\bigr)$ is a diagonal matrix containing the state-dependent distributions defined in \eqref{eqn:state_dep}.
% Time series of different defenders within the same play and of different plays are treated as independent, hence their log-likelihood contributions are summed to obtain the full likelihood of the training data. 

For practical implementation, we wrote a custom (negative joint log) likelihood function in \texttt{R}, using the function \texttt{forward\_(g)} 
% and other convenience functions 
from the \texttt{R} package \texttt{LaMa} \citep{koslikLaMa2025}, that allows evaluating the logarithm of \eqref{eqn:cond_lik} for all plays and all defenders, and thereby the computation of the term of interest $\log f_{\bm{\beta}, \sigma}(\bm{y} \vert \bm{u}, \bm{v}, \bm{w})$.

The parameters to be estimated are the two coefficients of the transition matrix, the standard deviation of the state-dependent distributions, and the three standard deviations of the random effects distributions. Due to the nature of the Laplace approximation --- performing an inner optimization over the random parameters for each outer iteration --- predicted random effects are also immediately available upon model fitting.

\subsubsection{Feature extraction from state decoding}\label{sec:summary}
The primary objective of our unsupervised approach is to extract informative features that enhance the supervised learning models. For this, we seek to decode the underlying guarding assignments throughout the motion phase based on the observed player trajectories. HMMs are particularly suitable for this task, as they exploit both the current observations --- such as the $\mathrm{y}$-coordinates of the individual defenders and all five offensive players --- and the temporal dependencies encoded through the transition probabilities. %By leveraging information from past and future time points, the model provides a probabilistically coherent inference of matchups between defensive and offensive players over time.
While global decoding methods (e.g., the Viterbi algorithm; \citealp{zucchini}) yield the single most likely state sequence, local decoding offers a richer quantification of uncertainty.
Thus, we compute the conditional probability of each latent guarding state at each time point $t$ given the trajectories of the defender and the offensive players in that play:
$$ \Pr(S_t = j \mid y_1, \dots, y_{T_m}) $$
where $y_1, \dots, y_{T_m}$ denotes the full sequence of positional observations up to time $T_m$, $m ~=~1, \ldots, M$. This approach retains the full probabilistic information and provides a categorical state distribution for each defender at each time point $t$. %These distributions can then be summarized into interpretable statistics—such as the mean or variance of guarding probabilities or the number of state switches—which serve as informative features in our post-motion supervised models.

%To obtain both the local state probabilities and the global state sequence, we used the functions `stateprobs()` and `viterbi()` that are also contained in the R package `LaMa`.

%As we aim to retrain the supervised mod-els described above, 

To incorporate the HMM results into the supervised models, we need to transform the state decoding time series as additional features. Integrating such derived information aligns with the principle of \textit{enhanced statistical learning}, in which additional, substantively meaningful information is leveraged to improve predictive accuracy \citep{felice2025boosting}. By enriching the feature space with information derived from the HMM results, the supervised models gain access to latent dynamics that would otherwise remain unobserved. However, including the decoded state probabilities directly into the supervised learning models is not straightforward, as the state sequences constitute a multi-dimensional time series: for each defender and each time point during the motion phases, we end up with five state probabilities, one for each potential offensive skill player to be guarded. % that varies across defenders, offensive players, and time. 
To render these time series suitable for the supervised
learning framework, we derive different informative metrics that capture the essential temporal
and probabilistic characteristics of the guarding dynamics.
In particular, for each defender, we first identify at each time point $t = 1, \ldots, T_m$ of play $m = 1, \ldots, M$ the most likely guarded offensive player given the full observation sequence,
\begin{equation}\label{eq:decod}    
\hat{S}_{t} = \argmax_{k \in \{1, \ldots, 5\}} \Pr(S_t = k \mid y_1, \dots, y_{T_m}).
\end{equation}
Subsequently, we count the number of state switches, i.e., instances where a defender’s most likely guarded player changes between two consecutive time points $t$ and $t+1$. From this sequence, we compute %several interpretable summary measures that describe the stability of the defensive assignments, such as 
the total number of switches within a play %their average across defenders, 
and the number of defenders who switch between offensive players at least once in the respective play. 
In addition to these discrete metrics, we compute a continuous statistic reflecting the uncertainty in the guarding assignments. Specifically, we calculate the (defensive) mean entropy across defenders $j = 1, \ldots, 5$ within each play $m = 1, \ldots, M$ (see \citealp{franks2015}) as
\[
H(m) = -\frac{1}{5} \sum_{j_m=1}^{5} \sum_{k=1}^{5} 
\left( 
\frac{1}{T_m} \sum_{t=1}^{T_m} \mathbbm{1}\left(\hat{S}_{t,j_m} = k\right)
\right)
\log\left(
\frac{1}{T_m} \sum_{t=1}^{T_m} \mathbbm{1}\left(\hat{S}_{t,j_m} = k\right)
\right),
\]
where $\hat{S}_{t,j_m}$ denotes the most likely guarded offensive player as defined in Eq.\ \ref{eq:decod} with the additional information of the $j$-th defender in play $m$.
%$$H(n) = - \frac{1}{5} \sum_{j, k=1}^{5} \left( \frac{1}{T_n} \sum_{t=1}^{T_n} \mathbbm{1}\left(\argmax_{i=1,\ldots,5} X_{t,i} = k\right) \log\left(\frac{1}{T_n} \sum_{t=1}^{T_n} \mathbbm{1}\left(\argmax_{i=1,\ldots,5} X_{t,i} = k\right)\right) \right)$$

Entropy, as a measure of uncertainty in a probability distribution, provides insight into the consistency of the defensive assignments: Lower entropy values indicate stable and predictable defensive structures, whereas higher values reflect increased variability and uncertainty in matchups. 

Finally, we also incorporate the estimated play-specific random effect --- which captures overarching tendencies to switch the guarding assignment in that respective play --- as an additional feature (see Eq.~\ref{eq:re}).

\subsection{Supervised learning}
Our supervised learning approach aims to predict the defensive coverage type --- man or zone --- based on several features. %As our dataset comprises only \( M = 3985 \) passing plays that include pre-snap motion, model complexity must be carefully controlled to avoid overfitting. %Since the main focus of this paper is the incorporation of features describing pre-snap motion, 
For this, we follow a three-step approach: First, we establish a baseline model for coverage type prediction using only the contextual and spatial pre-motion features derived in Section~\ref{sec:data}. This model serves two purposes. On the one hand, it provides a natural baseline for comparing models using motion information. On the other hand, this model is useful when analyzing the impact of using motion to detect the correct coverage scheme (see Section \ref{sec:team_analysis}). Second, we fit a model to an augmented set of features containing naive post-motion information derived in Section~\ref{sec:data}. Third, we fit a model using both the pre-motion and naive post-motion features and additionally the previously described HMM features. This enriched model therefore contains directly observable post-motion information but also latent assignment patterns inferred from the HMM, thereby capturing relevant dynamics of the defensive coverage.

To model the probability that a play \( i \in \{1, \dots, M\} \) corresponds to a man (\( y_i = 1 \)) or zone (\( y_i = 0 \)) coverage, we fit two complementary model classes: an elastic net logistic regression \citep{Zou05elastic} and gradient-boosted decision trees, implemented as an XGBoost model \citep{chen2016xgboost}. The former model is an interpretable model known to handle large feature spaces by regularization and implicit variable selection. However, its strong modeling assumptions, particularly the assumption of linear relationship between features and log odds of the outcome, may limit predictive performance. The latter is more flexible, allowing for non-linearities and interactions between features, at the cost of interpretability. Furthermore, tree boosting models typically require a large amount of data for proper hyperparameter tuning. In \texttt{R}, we use the \texttt{glmnet} \citep{friedman2021package} and the \texttt{xgboost} \citep{chen2019package} packages.

\subsection{Model evaluation and inference}
\label{sec:model_eval}
There are various objectives when evaluating the performance of our three-step supervised learning approach described in the previous section. One aim is to obtain a well-performing model for predicting the coverage type. That is, we want to compare the performance of the three models on a set of unseen data points. Commonly used loss metrics for a binary outcome are accuracy, i.e., the percentage of correctly classified observations; the area under the receiver operating characteristic curve (AUC), i.e., the area under the mapping between the true positive rate (TPR) and the false positive rate (FPR); %, or equivalently, the probability that the model ranks a randomly chosen positive instance higher than a randomly chosen negative instance, 
and the logloss, i.e., the (negative) log-likelihood value. For both the elastic net and the XGBoost model this coincides with a Bernoulli log-likelihood based on the predicted probability for each observation. 
The most popular loss metric is arguably the accuracy due to its simple interpretation. However, evaluating performance on accuracy can be misleading as it is dependent on the threshold selected for classification of the outcome \citep[see, e.g.,][]{tholke23acc}. Typically, for probabilistic predictions, as obtained by the two model classes considered, one uses the naive threshold of $0.5$. The AUC, on the other hand, is threshold independent as it compares TPR and FPR among a range of possible values. Hence, it is more suitable for binary classification, especially in imbalanced setups. From a mathematical perspective, the logloss is the most attractive loss metric due to being a proper scoring rule \citep{waghmare25scoring}, i.e., it is maximized in expectation by the true probabilities of each class. Hence, in this work, we mainly focus on the logloss to evaluate model performance, but for reasons of comparison, we also indicate the other loss metrics. 

While evaluating predictive performance on hold-out data allows for identifying the best-performing models, it is not suitable for inference on the features. More specifically, an important question is to identify which features are significantly associated with the outcome. %Or, in our case, we are interested in testing whether the 
In many fields, including sports analytics, the identification of influential factors is often approached by fitting machine learning models and examining traditional variable importance measures \citep{bajons2025mlsi}. However, variable importance measures differ by the regression technique used, hence do not allow for comparison between regression models \citep{williamson20varimp}. Additionally, there are various ways to compute variable importance measures for tree ensembles as used in this work, each having its own advantages and disadvantages \citep{Strobl08varimp}. A recently popularized alternative are non-parametric conditional independence tests. As opposed to traditional variable significance tests, which assume a parametric relationship between outcome and features, these tests determine which features are significantly associated with an outcome with considerably relaxed assumptions, generally relying on modern machine learning algorithms. In this work, we consider a popular family of such tests commonly referred to as covariance measure tests \citep[COMETs, see, e.g.,][]{kook24comets}, and specifically, the Generalized Covariance Measure (GCM) test \citep{Shah20gcm}. COMETs test for conditional independence of an outcome variable $Y$ and a (possibly vector-valued) feature $X$ given additional features $Z$, without imposing strong (parametric) assumptions on the (conditional) distribution of $Y$.
More formally, the GCM test assesses the conditional independence of $Y$ and $X$ given $Z$, denoted by $Y \perp\!\!\!\perp X \mid Z$, by targeting the 
\begin{equation}\label{eq:GCM}
\operatorname{GCM} \coloneqq \mathbb{E}[\operatorname{Cov}(Y,X \mid Z)] =\mathbb{E}[(Y -
\mathbb{E}[Y | Z])(X - \mathbb{E}[X | Z])],
\end{equation}
and using the fact that a necessary condition for $Y \perp\!\!\!\perp X \mid Z$ is that $\mathbb{E}[\operatorname{Cov}(Y,X \mid Z)] = 0$. To obtain an empirical estimate of the GCM, one therefore has to learn two regression functions $h(Z) \coloneqq \mathbb{E}[Y | Z]$ and $f(Z) \coloneqq \mathbb{E}[X | Z]$. The test remains valid under mild assumptions, which are typically met when these functions are estimated by suitable machine learning methods \citep{Shah20gcm}, and thus remains applicable even without imposing strong modeling assumptions.
% Hence, these tests can be used even when modeling the outcome with modern machine learning algorithms. 
Aside from its statistical merit, the GCM test is also appealing from an interpretational viewpoint. \citet{bajons2025rGAX} show that for a binary outcome, as in our case, the GCM test can be related to a parameter estimate in semi-parametric partially linear logistic regression model.
More precisely, let $(Y, X, Z)$, where $Y \in \{0, 1\}$, $X \in \mathbb{R}^{d_X}$, $Z \in \mathbb{R}^{d_Z}$, be a random vector in which $Y$ is governed by a partially linear regression model. That is, $Y \given X, Z \sim \operatorname{Ber}(\pi(X,Z))$
follows a Bernoulli distribution with $\pi(X,Z) = P(Y=1 \mid X,Z) = \mathbb{E}[Y \given
X,Z]$, and 
\begin{align}
\label{eq:pllm_lo}
\begin{aligned}
\log\left(\frac{\pi(X,Z)}{1-\pi(X,Z)}\right) =
X\beta + g(Z),
\end{aligned}
\end{align}
with some arbitrary measurable function $g$. Then, \citet{bajons2025rGAX} show that the GCM test can be used to test hypotheses of the form $H_0: \beta = 0$ as well as $H_0 \ge 0$ (respectively $H_0 \le 0$) in the PLLM.  
That is, using the GCM test, we can not only identify which feature is significantly associated with the outcome, but also interpret the direction that a feature influences it in the context of the model from Eq.~\ref{eq:pllm_lo}. In our case, we use the GCM test implemented in \texttt{R} via the \texttt{comets} package (\citealp{kook24comets}) to assess whether HMM-derived features significantly influence man-versus-zone predictions after conditioning on all other variables, including naive post-motion features. Additionally, we interpret the results of the tests within the partially linear logistic regression model of Eq.~\ref{eq:pllm_lo}.

\section{Results}\label{sec:results}
In this section, we first present results from the hidden Markov model (HMM), including estimated transition dynamics, random effects, and decoded state probabilities that capture latent guarding assignments. We then report results from supervised learning models predicting defensive coverage schemes. Next, we evaluate whether the additional HMM-derived features yield statistically significant improvements in predictive performance. Finally, we demonstrate the practical utility of our framework through team-level analyses.

\subsection{Unsupervised learning}
Before fitting the full HMM incorporating random effects in its state process, we first estimated a simpler version with a homogeneous Markov chain to determine the optimal time lag $l$ corresponding to defenders’ reaction times (see Eq. \ref{eq:lag}). The figure in Section~2 of the Supplementary Material indicates that $l = 4$ provides the best fit according to AIC values. This aligns closely with previously reported reaction times in soccer (\citealp{wu2023evaluation}). Therefore, we use $l = 4$ when fitting our full model.

The most intriguing results --- apart from the decoded state sequences --- relate to the estimated transition probability matrix, which includes the random effects. Table~\ref{tab:position_effects} summarizes the predicted random effects across positions. These effects quantify how state-switching probabilities vary by role. Smaller values correspond to a higher likelihood of remaining in the current state, i.e., continuing to guard the same offensive player at subsequent time steps. Consistent with visual observation, cornerbacks exhibit strong tendencies to follow receivers across plays, whereas linebackers and safeties more frequently pass off coverage assignments.

\begin{table}[b!]
\centering
\caption{Predicted random effects separated for each defensive position.}
\label{tab:position_effects}
\begin{tabular}{|l|c|}
\hline
Position & Predicted Effect \\
\hline
Inside Linebacker & $\ \ \  0.007$\\
Middle Linebacker & $\ \ \    0.006$\\
Outside Linebacker & $\ \ \ 0.006$\\
Cornerback & $-0.178$ \\
Strong Safety & $-0.005$\\
Free Safety & $\ \ \ 0.175$\\
\hline
\end{tabular}
\end{table}

Similarly, Section~4 in the Supplementary Material displays the predicted random effects for all defensive teams. These estimates reflect structural differences in defensive schemes between teams. 

The position-, team-, and play-level random effects are incorporated into the overall state-switching dynamics of the HMM. Since the underlying Markov chain is modeled in a non-homogeneous manner, it is not possible to present a single transition probability matrix. Nevertheless, to provide intuition regarding the state transitions, we report the estimated transition probability matrix (t.p.m.) for Marco Wilson, a cornerback of the Arizona Cardinals, in the play shown in Figure~\ref{fig:prepro}. Specifically, this t.p.m.\ corresponds to the moment when Mecole Hardman (MH) --- the player Marco Wilson is covering --- crosses running back Jerick McKinnon (JMK). This example demonstrates the strength of our approach: although Wilson is momentarily as close to McKinnon as to Hardman, his estimated tendency to stay with Hardman remains high (see $\hat{\gamma}_{11}$), reflecting realistic persistence in man‑to‑man coverage behavior.

$$
\renewcommand{\arraystretch}{1.2}
\setlength{\arraycolsep}{4pt}
\begin{array}{c|c}
    &
\hspace{-0.8em} % etwas nach links schieben
    \begin{array}{ccccc}
        \ \ \ \ \ \ \text{MH} \ \ \ \ & \text{TK} \ \ \ & \text{JMK} \ \ \ & \text{JSS} \ \  & \text{MVS}\ \ 
    \end{array}
    \hspace{-1em} % etwas nach rechts schieben
    \\[4pt] \hline
    \begin{array}{c}
        \text{MH} \\[1pt] 
        \text{TK} \\[1pt]
        \text{JMK} \\[1pt]
        \text{JSS} \\[1pt]
        \text{MVS} 
    \end{array}
    &
    \left(
      \begin{array}{ccccc}
      0.983 & 0.002 & 0.015 & 0.000 & 0.000 \\
      0.002 & 0.997 & 0.002 & 0.000 & 0.000 \\
      0.015 & 0.002 & 0.983 & 0.000 & 0.000 \\
      0.000 & 0.000 & 0.000 & 1.000 & 0.000 \\ 
      0.000 & 0.000 & 0.000 & 0.000 & 1.000
      \end{array}
    \right)
\end{array}
$$

Overall, estimated transition probabilities indicate largely stable and persistent latent-state dynamics. Occasionally elevated state-switching probabilities (such as $\hat{\gamma}_{13}$) occur when defenders move toward offensive players occupying similar horizontal zones (e.g., JMK). In contrast, transitions to more distant offensive targets (Travis Kelce, TK; Juju Smith-Schuster, JSS; or Marquez Valdes-Scantling, MVS) are seldom observed.

\begin{figure}
    \centering
    \includegraphics[width=\linewidth]{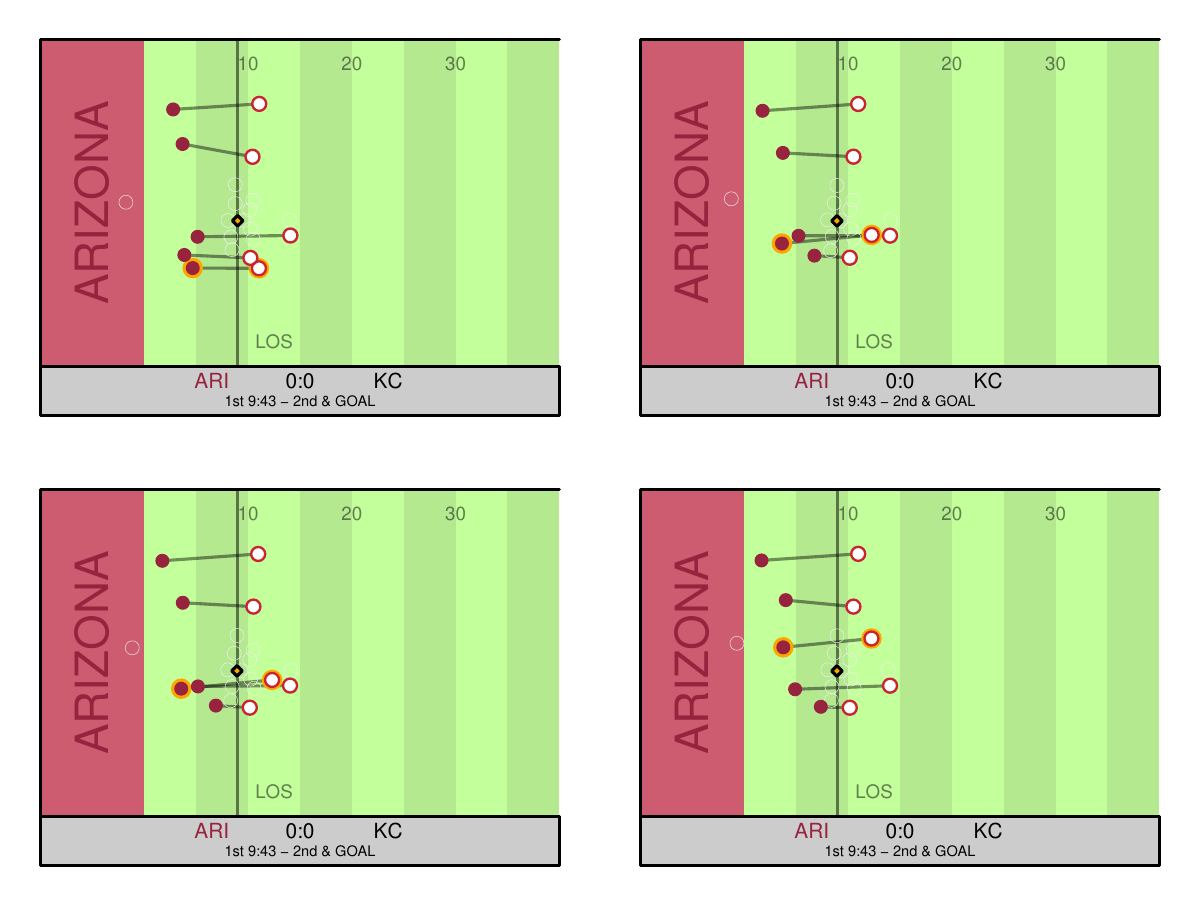}
    \caption{This figure illustrates how guarding assignments evolve during an example play based on decoded state sequences from our HMM analysis. Lines between offensive and defensive players represent inferred matchups over time. The top-left panel shows initial formations at the line of scrimmage; subsequent panels depict intermediate stages culminating in the bottom-right frame immediately before the snap. Notably, when Mecole Hardman (MH) crosses paths with running back Jerick McKinnon (JMK), Marco Wilson’s coverage remains firmly maintained on Hardman.}
    \label{fig:snap}
\end{figure}

\begin{table}[!t]
\caption{Summary statistics of the features that were extracted from the multi-dimensional time series of state decodings.}
\label{tab:state_decod}
\centering
\begin{tabular}[t]{l|cccc}
\hline
statistic & Total switches & \# switching defenders & avg.\ entropy & RE/play\\
\hline
Min. & 0.00 & 0.00 & 0.00 & -0.91\\
1st Qu. & 0.00 & 0.00 & 0.00 & -0.18\\
Median & 1.00 & 1.00 & 0.13 & -0.03\\
Mean & 1.82 & 1.45 & 0.17 & 0.00\\
3rd Qu. & 3.00 & 2.00 & 0.27 & 0.16\\
Max. & 21.00 & 5.00 & 1.01 & 1.34\\
\hline
\end{tabular}
\end{table}

While these results offer valuable insights into latent state-switching dynamics, our main focus lies in the decoded state sequences and their translation into the summary statistics described above. As an example of a decoded time series, Figure~\ref{fig:snap} illustrates how the HMM effectively captures Marco Wilson's guarding assignment\footnote{A complete animation of the sequence is provided as supplementary material.}. Interestingly, the unique strength of HMMs becomes apparent when Wilson reaches the same $\mathrm{y}$-coordinate as running back Jerick McKinnon. At this moment, the HMM consistently and accurately maintains the correct coverage assignment throughout the motion as a consequence of its temporal persistence. In contrast, classical clustering algorithms based solely on spatial proximity, ignoring the temporal dependencies, would likely suggest a change in guarding assignment, thereby introducing noise. 

We transformed the decoded state sequences into the metrics described in Section~\ref{sec:summary}. Table~\ref{tab:state_decod} summarizes key descriptive statistics such as the minimum, maximum and mean of the number of state switches, the number of defenders who change their guarding assignment throughout the motion phase, the defensive mean entropy across defenders and the predicted random effect per play.

\subsection{Supervised learning}

\begin{figure}
    \centering
    \includegraphics[width=\linewidth]{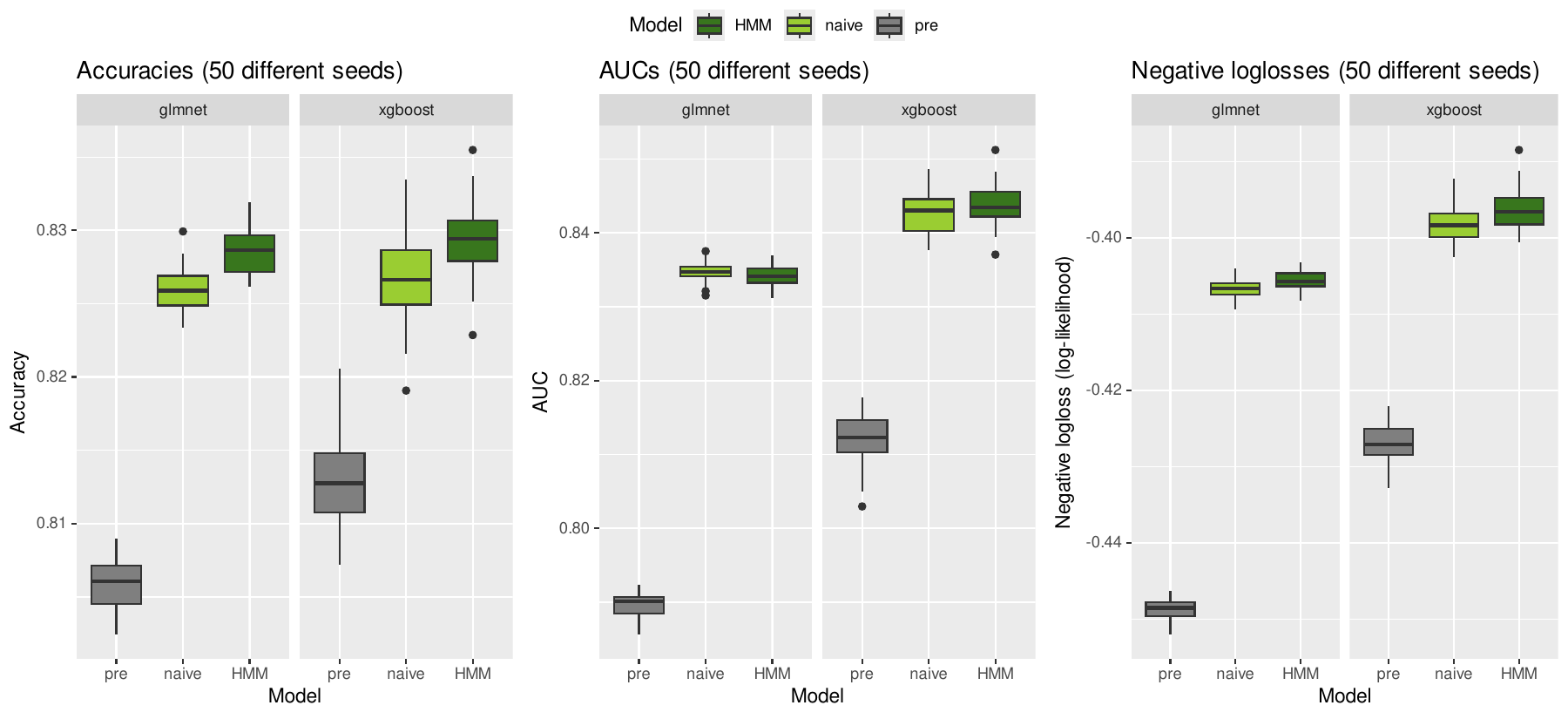}
    \caption{Boxplots of the different loss metrics (left subpanel: accuracies; middle: AUC; right: negative logloss) obtained from 50 different runs of the cross fitting procedure. Each panel contains the results from the elastic net (\texttt{glmnet}) on the left and the XGBoost model, on the right.}
    \label{fig:loss_metrics}
\end{figure}

As described in Section \ref{sec:model_eval}, we first compare the performance of the baseline model using only pre-motion features with the model using additional naive post-motion features, % (denoted by ``naive''), 
and the model using all previous features and the HMM-derived features. % (denoted by \textit{HMM}). 
Since we are working with a small data set, we implement a cross-fitting strategy to obtain out-of-sample predictions for each observation. Specifically, we randomly split the full data into five folds. For each fold, we train a model on all data except the data contained in the selected fold.  Both model classes --- the elastic‑net and XGBoost models --- require specification of hyperparameters; therefore, within each training iteration we perform 5‑fold cross-validation on the remaining four folds to tune these parameters. More precisely, we use a pre-defined grid to tune \( \lambda \) and $\alpha$ for the elastic net, and tree depth, learning rate, and number of boosting rounds for XGBoost. The tuned models are then used to generate predictions for all observations in the left-out fold. Repeating this procedure yields out-of-sample predictions for every plan in the data set. Hereby, we can compare our three models using the different loss metrics (accuracy, AUC, and logloss) on unseen data. However, because cross‑fitting involves random partitioning into five folds and further random splits during inner cross-validation for hyperparameter tuning, estimates of out‑of‑sample losses may vary depending on these random draws. To address this source of uncertainty, we repeat the entire cross-fitting routine 50 times and report resulting distributions of loss metrics rather than a single value from one random split.

%For both models, we conduct 10-fold cross-validation over a pre-defined hyperparameter grid to identify optimal values of \( \lambda \) and $\alpha$ (for the logistic elastic net), as well as tree depth, learning rate, and number of boosting rounds (for the XGBoost model).

Figure \ref{fig:loss_metrics} displays the results of the cross-fitting loss evaluation procedure. Across all three loss metrics, we observe that the more flexible XGBoost models consistently outperform the elastic net models (referred to as \texttt{glmnet}). Furthermore, the pre-motion models perform worst, suggesting that even a naive incorporation of motion information substantially improves prediction of the defensive coverage scheme. While this is an expected finding, we also observe that including HMM-derived features enhances predictive performance. These results indicate that features derived from decoded state sequences of the HMM capture latent information beyond simple motion characteristics and provide additional insight for predicting overall coverage schemes.

\begin{figure}
    \centering
    \includegraphics[width=0.5\linewidth]{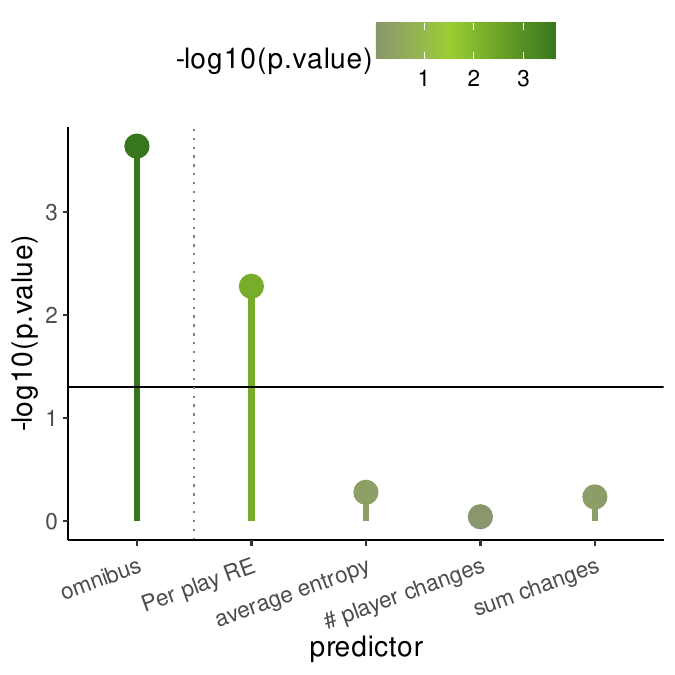}
    \caption{$p$-values on a $-\log_{10}$-scale of the GCM test for conditional independence of HMM features and coverage outcome given pre- and naive post motion features as well as all other HMM features for the individual tests. The solid horizontal line indicated the 5\%-level. The dashed vertical line separates the omnibus test, i.e., a test on all features simultaneously from the tests on specific features.}
    \label{fig:sig1}
\end{figure}

While Figure~\ref{fig:loss_metrics} already suggests improved predictive performance from incorporating HMM-derived features, further evidence of their usefulness comes from the GCM test described in Section~\ref{sec:model_eval}. This test allows us to statistically assess whether these features add significant predictive power. For this analysis, we used the best-performing model, i.e., the XGBoost model, as the regression method for the estimation of $h(Z)$ and $f(Z)$ (see Section~ \ref{sec:model_eval}). More specifically, the GCM test can be applied in several ways. Let $X \in \mathbb{R}^4$ denote the set of HMM features; we first test the null hypothesis $Y \perp\!\!\!\perp X \mid Z$, i.e., we perform an omnibus test, which assesses whether at least one of the HMM-derived features is significantly associated with the defensive coverage outcome $Y$. The result of this test is shown in Figure \ref{fig:sig1}. The test is highly significant at every conventional level ($p$-value $\approx 0.0002$), strongly indicating that the HMM-derived features aid in predicting the defensive coverage scheme even in the presence of pre- and naive post-motion features. Next, we test for each feature in $X$ individually, i.e., perform a test of the form $H_0^j: Y \perp\!\!\!\perp X_j \mid (Z,X_{-j})$, $j = 1,\dots,4$, where $-j$ denotes the set of features when removing feature $j$. Figure~\ref{fig:sig1} demonstrates that only the feature \textit{Per play RE} (representing estimated play-specific random effects influencing the state transition probability) is significant at 5\% and 1\% level ($p$-value $\approx 0.005$). For the other features, the test is not rejected at conventional levels. While this result may seem surprising, it is much likely explained by strong correlation among the HMM features. 

\begin{figure}
    \centering
    \includegraphics[width=0.5\linewidth]{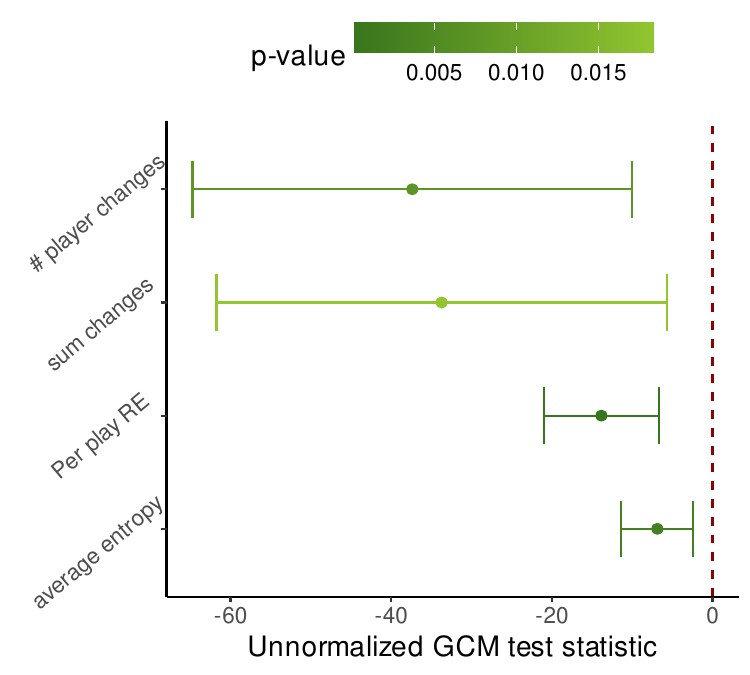}
    \caption{The unnormalized test statistic of the GCM test with 95\% confidence interval. The vertical dashed line corresponds to the zero value.}
    \label{fig:sig2}
\end{figure}

To gain deeper insight into the individual HMM-derived features, we analyze them within the framework of the partially linear logistic regression model defined in Eq.~\ref{eq:pllm_lo}. To this end, we consider each feature $X_j$ individually. More precisely, we test the hypotheses $H_0^j: Y \perp\!\!\!\perp X_j \mid (Z,X_{-j})$, $j = 1,\dots,4$, which corresponds to testing $H_0: \beta = 0$ in a partially linear logistic regression model for $(Y,X_j,Z)$. Figure~\ref{fig:sig1} shows the resulting GCM test statistics and confidence intervals for each feature. First, we observe that all features are significant (the confidence intervals do not include zero). Second, the test statistics are negative for all HMM-derived features (see Figure~\ref{fig:sig2}). This implies that the coefficient $\beta$ in Eq.~\ref{eq:pllm_lo} is significantly negative, indicating a negative effect of all features on the probability of man coverage. This finding aligns with our intuition. Lower values of the features describing the number of switches and the total count of assignment changes indicate lower variability in latent guarding patterns, typically a characteristic of man coverage. The average entropy is somewhat more complex to interpret but conceptually similar to these two metrics; hence its significantly negative impact on coverage outcome appears equally reasonable. The estimated random effect from the HMM model that we include as a feature is different in nature to the previous three. It influences transition probabilities of the latent Markov chain (see Section \ref{sec:MC_dets}). A higher estimate corresponds to an increased probability of switching states, which in turn represents changes in latent guarding assignments. Hence, the strongly negative impact on the probability of man coverage is also reasonable in this case.

\subsection{Team analysis}
\label{sec:team_analysis}

\begin{figure}
    \centering
    \includegraphics[width=\linewidth]{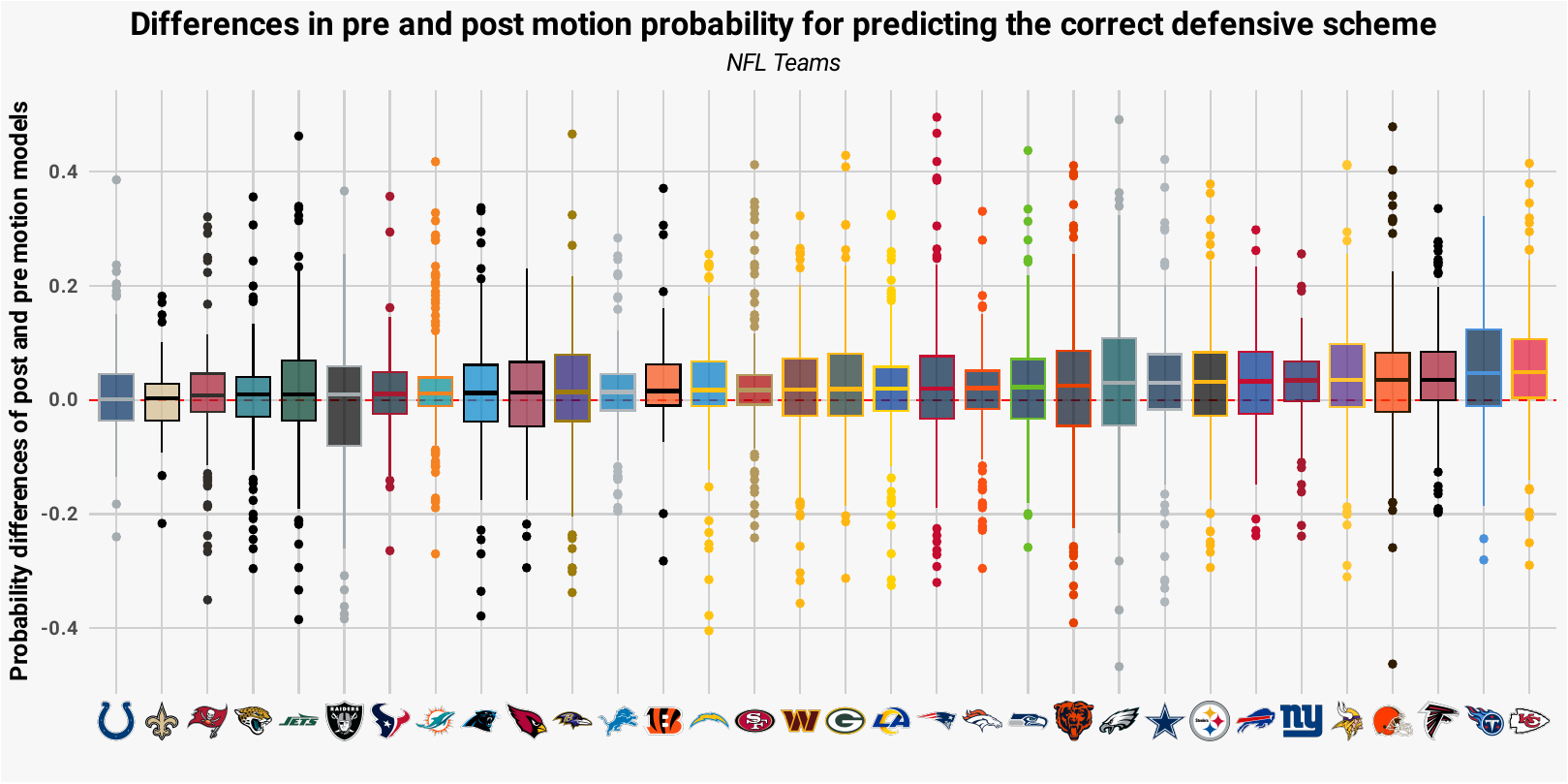}
    \caption{Boxplots of the differences in the prediction of the correct defensive scheme between a model using motion features and a model using only pre motion features for each play of each team in the NFL. Teams are ordered by medians of the boxplots.}
    \label{fig:teamdiff}
\end{figure}

While this study primarily demonstrates how HMM-derived features enhance play-level predictions of defensive coverage schemes, the final model can also be used to evaluate teams’ ability to use pre‑snap motion to identify the correct coverage, assuming that their visual identification aligns with our statistical forecasts.  

\begin{figure}
    \centering
    \includegraphics[width=0.65\linewidth]{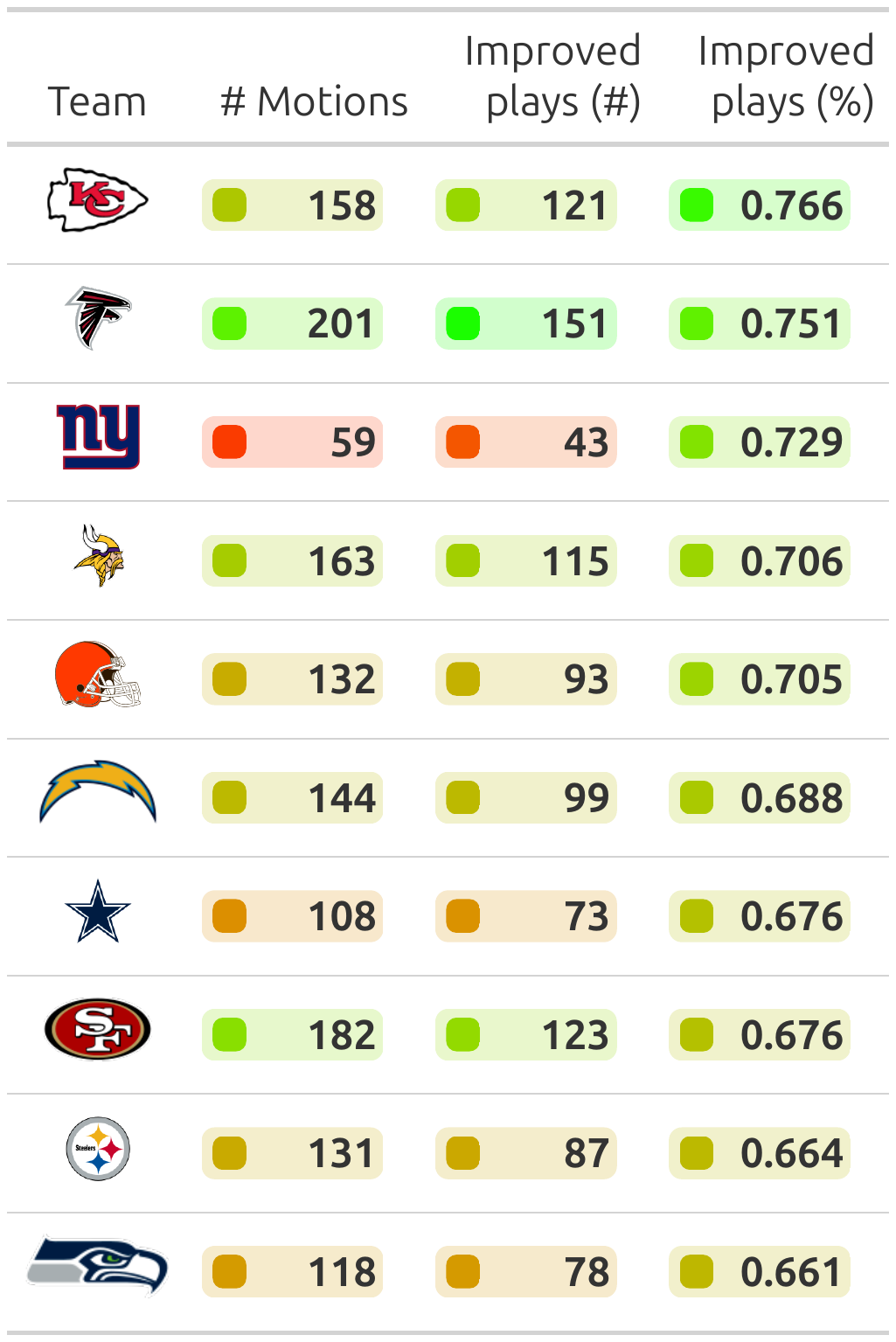}
    \caption{A table indicating the number of motion plays, the number of plays where the motion improved the probability of predicting the correct outcome, and proportion of motion plays where probability of correct outcome was improved for the top ten NFL teams with respect to improvement percentage.}
    \label{fig:team_tab}
\end{figure}

To assess how effectively teams utilize pre-snap motion to recognize defensive coverage schemes, we compare predictions from two of the three supervised models: the pre-motion XGBoost model and the post-motion XGBoost model including HMM features. For estimation, we adopt an out-of-sample prediction strategy to avoid data leakage between training and evaluation phases. Specifically, for each team under consideration, all plays belonging to that team are excluded during model training and hyperparameter tuning. Afterwards, the two models are trained on data from all remaining teams using cross-validation procedures described earlier. Predictions for the omitted team’s plays are then generated using these fitted models. Although this leave-one-team-out procedure may not exploit intra-team characteristics optimally due to data limitations, it ensures unbiased performance assessment across teams. We then evaluate the corresponding predicted probabilities of correct classification obtained from both models. Conceptually, if pre-snap motion assists offenses in recognizing defensive structures, predictions from the model including motion should yield higher probabilities than those from the model without motion.
Figure~\ref{fig:teamdiff} visualizes these differences using boxplots for each team, ordered by their median improvement values. The results reveal considerable variation across teams but generally indicate positive median differences, suggesting that most teams benefit from motion-based information when anticipating coverages. Teams such as Kansas City Chiefs, Tennessee Titans, and Atlanta Falcons exhibit the highest median improvements and thus appear particularly effective at leveraging pre-snap motion.

Complementary insights are provided by Figure~\ref{fig:team_tab}, which lists the ten teams with the largest proportion of plays where post-motion predictions improved relative to pre-motion ones. Again, Kansas City --- the team that has been used as the running example in this study --- emerges as the most effective team overall, while Atlanta and San Francisco --- both known for their frequent use of pre-snap motion under Arthur Smith and Kyle Shanahan --- also rank among the top performers. Interestingly, the Giants demonstrate high effectiveness despite employing relatively little motion, indicating efficient recognition of coverages even with limited movement.

%As shown in the Figure~\ref{fig:teamdiff}, teams that are well-known for extensive use of pre-snap motion --- such as the San Francisco 49ers--- are also among the ones that most frequently increased the likelihood of correctly decoding the defensive coverage. While some teams --- such as the NY Giants --- succeed in accurately identifying coverages despite employing limited motion, the general trend indicates that teams utilizing extensive motion are more effective at leveraging it to recognize defensive schemes.

\section{Discussion}\label{sec:discussion}

This study combined supervised and unsupervised learning techniques to improve the prediction of defensive schemes in American football based on pre‑snap player movements. By augmenting baseline models with HMM‑derived features, we observed consistent improvements in predictive performance relative to approaches relying solely on static pre‑snap and naive post-motion features. Moreover, we showed using a non‑parametric test that HMM-based features significantly contribute to predicting coverage outcomes. These findings suggest that latent‑state representations can effectively capture underlying behavioral dynamics that are otherwise obscured.

Despite these encouraging results, several limitations should be noted. The available dataset (i.e., the number of plays) was quite small, constraining both model complexity and generalizability. Consequently, our post-motion models may not fully exploit the richness of the information derived from the HMM. Nevertheless, a key advantage of our framework lies in its modularity: the HMM component can be seamlessly integrated with alternative modeling frameworks such as neural networks (\citealp{Song2023}). A major advantage would be that those architectures are potentially capable of dealing with the raw decoded latent state sequence rather than relying solely on summary statistics which we were not able to do due to the limited training data (\citealp{dempster2020rocket}). 

Another limitation of our approach lies in the granularity of the defensive scheme classification. Our model distinguishes only between man and zone coverage, which represents a coarse abstraction of real-world defensive strategies. In practice, each category encompasses multiple subtypes, such as Cover-1, Cover-2, Cover-3, or hybrid schemes. Extending the framework to capture these finer-grained latent structures would allow for a more nuanced understanding of defensive behavior. In this context, HMMs could be used to model not only the defensive responsibilities but also switches between man-coverage and specific areas of the field directly (see \citealp{groom} for a similar approach in soccer); potentially within a hierarchical HMM structure.

Finally, albeit we include a play-specific random effect that accounts for similar behavior of the defenders, assuming independence between defenders is a rather strong assumption. To relax this, future research could exploit recent advancements that suggest how to model multivariate state-dependent distributions non-parametrically within an HMM (\citealp{koslik2025tensor, michels2025nonparametric}). In this way, dependencies between defender can be modeled, which might lead to better insights into the coordinated defensive behavior and, finally, more informative features for the overall supervised model.

Overall, our results demonstrate that the integration of a hidden Markov model offers a promising avenue for capturing defensive tasks in American football. With expanded datasets, this line of research could substantially advance quantitative understanding of strategic decision-making in American football, but also beyond that, in team sports.

\section{Code availability}

All code can be found on GitHub (\url{https://github.com/janolefi/BDB-2025/}).

%%%%%%%%%%%%%%%%%%%%%%%%%%%%%%%%%%%%%%%%%%%%%%
%% Appendix---Please move all appendices to %%
%% a Supplementary file.                    %%
%%%%%%%%%%%%%%%%%%%%%%%%%%%%%%%%%%%%%%%%%%%%%%
%% Support information, if any,             %%
%% should be provided in the                %%
%% Acknowledgements section.                %%
%%%%%%%%%%%%%%%%%%%%%%%%%%%%%%%%%%%%%%%%%%%%%%
\section*{Acknowledgments}
The authors would like to thank the organizers of the NFL Big Data Bowl 2025 for setting up this competition and providing access to the data.
Moreover, the authors gratefully acknowledge the computing time provided on the Linux HPC cluster at Technical University Dortmund (LiDO3), partially funded in the course of the Large-Scale Equipment Initiative by the Deutsche Forschungsgemeinschaft (DFG, German Research Foundation) as project 271512359.
\bibliographystyle{apalike} % Style BST file
\bibliography{refs}       % Bibliography file (usually '*.bib')

@article{franks2015,
	title        = {Characterizing the spatial structure of defensive skill in professional basketball},
	author       = {Franks, Alexander and Miller, Andrew and Bornn, Luke and Goldsberry, Kirk},
	year         = 2015,
	journal      = {The Annals of Applied Statistics},
	publisher    = {Institute of Mathematical Statistics},
	volume       = 9,
	number       = 1,
	pages        = {94--121}
}

@book{zucchini,
	title        = {Hidden {M}arkov {M}odels for {T}ime {S}eries: {A}n {I}ntroduction {U}sing {R}},
	author       = {Zucchini, Walter and MacDonald, Iain L. and Langrock, Roland},
	year         = 2016,
	publisher    = {Chapman and Hall/CRC},
	doi          = {10.1201/b20790},
address = {New York}
}

@Manual{koslikLaMa2025,
  title = {{L}a{M}a: {F}ast {N}umerical {M}aximum {L}ikelihood {E}stimation for {L}atent {M}arkov {M}odels},
  author = {Jan-Ole Koslik},
  year = {2025},
  note = {{R} package version 2.0.6},
  url = {https://CRAN.R-project.org/package=LaMa},
}

@article{michels2024combination,
  title={On the combination of data smoothing and {M}arkov-switching models},
  author={Michels, Rouven and Koslik, Jan-Ole},
  journal={Journal of the Royal Statistical Society Series C: Applied Statistics},
  volume={73},
  number={3},
  pages={557--560},
  year={2024},
  publisher={Oxford University Press UK},
  note={\url{https://doi.org/10.1093/jrsssc/qlad110}}
}

@article{adam2024markov,
  title={Markov-switching decision trees},
  author={Adam, Timo and {\"O}tting, Marius and Michels, Rouven},
  journal={AStA Advances in Statistical Analysis},
  volume={108},
  number={2},
  pages={461--476},
  year={2024},
  publisher={Springer},
  note={\url{https://doi.org/10.1007/s10182-024-00501-6}}
}

@article{michels2025nonparametric,
  title={{Nonparametric estimation of bivariate hidden {M}arkov models using tensor-product B-splines}},
  author={Michels, Rouven and Langrock, Roland},
  journal={Statistical Modelling},
  pages={1471082X251335431},
  year={2025},
  publisher={SAGE Publications Sage India: New Delhi, India},
  note={\url{https://doi.org/10.1177/1471082X251335431}}
}

@article{dutta2020unsupervised,
  title={Unsupervised methods for identifying pass coverage among defensive backs with {NFL} player tracking data},
  author={Dutta, Rishav and Yurko, Ronald and Ventura, Samuel L},
  journal={Journal of Quantitative Analysis in Sports},
  volume={16},
  number={2},
  pages={143--161},
  year={2020},
  publisher={De Gruyter}
}

@article{groom,
  title={A Machine Learning Framework for Off Ball Defensive Role and Performance Evaluation in Football},
  author={Groom, Sean and Wang, Shuo and Belo, Francisco and Rice, Axl and Anderson, Liam},
  journal={arXiv preprint arXiv:2601.00748},
  year={2026}
}

@Manual{rtmb,
    title = {RTMB: 'R' Bindings for 'TMB'},
    author = {Kasper Kristensen},
    year = {2025},
    note = {R package version 1.7},
    url = {https://github.com/kaskr/rtmb},
  }

@article{bajons2025pep,
  title={{PEP}: a tackle value measuring the prevention of expected points},
  author={Bajons, Robert and Koslik, Jan-Ole and Michels, Rouven and {\"O}tting, Marius},
  journal={Journal of Quantitative Analysis in Sports},
  year={2025},
  publisher={De Gruyter},
    note={\url{https://doi.org/10.1515/jqas-2024-0099}}
}

@article{albert93,
	title        = {A Statistical Analysis of Hitting Streaks in Baseball: comment},
	author       = {Jim Albert},
	year         = 1993,
	journal      = {Journal of the American Statistical Association},
	publisher    = {[American Statistical Association, Taylor & Francis, Ltd.]},
	volume       = 88,
	number       = 424,
	pages        = {1184--1188},
	urldate      = {2023-11-02}
}

@article{michels2023bettors,
	title        = {Bettors' reaction to match dynamics: evidence from in-game betting},
	author       = {Michels, Rouven and {\"O}tting, Marius and Langrock, Roland},
	year         = 2023,
	journal      = {European Journal of Operational Research},
	publisher    = {Elsevier},
	volume       = 310,
	number       = 3,
	pages        = {1118--1127},
    note={\url{https://doi.org/10.1016/j.ejor.2023.04.006}}
}

@article{otting2023football,
	title        = {Football tracking data: a copula-based hidden {M}arkov model for classification of tactics in football},
	author       = {{\"O}tting, Marius and Karlis, Dimitris},
	year         = 2023,
	journal      = {Annals of Operations Research},
	publisher    = {Springer},
	volume       = 325,
	number       = 1,
	pages        = {167--183}
}

@article{felice2025boosting,
  title={Boosting any learning algorithm with Statistically Enhanced Learning},
  author={Felice, Florian and Ley, Christophe and Bordas, St{\'e}phane PA and Groll, Andreas},
  journal={Scientific Reports},
  volume={15},
  number={1},
  pages={1605},
  year={2025},
  publisher={Nature Publishing Group UK London}
}

@article{winkelmann2025momentum,
  title={Momentum effects in team sports: analyzing the interplay between offense and defense in the {NBA}},
  author={Winkelmann, David and Michels, Rouven},
  journal={The American Statistician},
  year={2025},
  publisher={Taylor \& Francis},
  note={\url{https://doi.org/10.1080/00031305.2025.2595980}}
}

@Article{Song2023,
 author = {Huan Song and Mohamad Al Jazaery and Haibo Ding and Lin Lee Cheong and Jonathan Jung and Mike Band and Michael Chi and Tom Bliss},
 title = {Explainable defense coverage classification in NFL games using deep neural networks},
 year = {2023},
 url = {https://www.amazon.science/publications/explainable-defense-coverage-classification-in-nfl-games-using-deep-neural-networks},
}

@article{sandholtz2024learning,
  title={Learning risk preferences in {M}arkov decision processes: An application to the fourth down decision in the {N}ational {F}ootball {L}eague},
  author={Sandholtz, Nathan and Wu, Lucas and Puterman, Martin and Chan, Timothy CY},
  journal={The Annals of Applied Statistics},
  volume={18},
  number={4},
  pages={3205--3228},
  year={2024},
  publisher={Institute of Mathematical Statistics}
}

@article{yurko,
	title        = {Going deep: models for continuous-time within-play valuation of game outcomes in {A}merican football with tracking data},
	author       = {Ronald Yurko and Francesca Matano and Lee F. Richardson and Nicholas Granered and Taylor Pospisil and Konstantinos Pelechrinis and Samuel L. Ventura},
	year         = 2020,
	journal      = {Journal of Quantitative Analysis in Sports},
	volume       = 16,
	number       = 2,
	pages        = {163--182},
	doi          = {10.1515/jqas-2019-0056},
	lastchecked  = {2022-10-14}
}

@article{brill2025analytics,
  title={Analytics, have some humility: a statistical view of fourth-down decision making},
  author={Brill, Ryan S and Yurko, Ronald and Wyner, Abraham J},
  journal={The American Statistician},
  year={2025},
  publisher={Taylor \& Francis},
note={in press}
}

@article{nguyen,
  title={Here comes the strain: Analyzing defensive pass rush in American football with player tracking data},
  author={Nguyen, Quang and Yurko, Ronald and Matthews, Gregory J},
  journal={The American Statistician},
  volume={78},
  number={2},
  pages={199--208},
  year={2024},
  publisher={Taylor \& Francis}
}

@article{kristensen2016tmb,
  title={{TMB}: automatic differentiation and {L}aplace approximation},
  author={Kristensen, Kasper and Nielsen, Anders and Berg, Casper W and Skaug, Hans and Bell, Bradley M},
  journal={Journal of Statistical Software},
  volume={70},
  pages={1--21},
  year={2016}
}

@article{mews2025build,
  title={How to build your latent {M}arkov model: {T}he role of time and space},
  author={Mews, Sina and Koslik, Jan-Ole and Langrock, Roland},
  journal={Statistical {M}odelling},
note = {in press},
  doi={10.1177/1471082X251355681},
  publisher={SAGE Publications Sage India: New Delhi, India},
  year={2025}
}

@article{lystig2002exact,
  title={Exact computation of the observed information matrix for hidden {M}arkov models},
  author={Lystig, Theodore C and Hughes, James P},
  journal={Journal of {C}omputational and {G}raphical {S}tatistics},
  volume={11},
  number={3},
  pages={678--689},
  year={2002},
  publisher={Taylor \& Francis}
}

@article{gay1990usage,
  title={Usage summary for selected optimization routines},
  author={Gay, David M},
  journal={Computing {S}cience {T}echnical {R}eport},
  volume={153},
  number={153},
  pages={1--21},
  year={1990}
}

@article{waghmare25scoring,
   author = "Waghmare, Kartik and Ziegel, Johanna",
   title = "Proper Scoring Rules for Estimation and Forecast Evaluation",
   journal = "Annual Review of Statistics and Its Application",
   year = "2025",
   publisher = "Annual Reviews",
   doi = "https://doi.org/10.1146/annurev-statistics-042424-050626",
   note = "in press"
}

@article{tholke23acc,
title = {Class imbalance should not throw you off balance: Choosing the right classifiers and performance metrics for brain decoding with imbalanced data},
journal = {NeuroImage},
volume = {277},
pages = {120253},
year = {2023},
doi = {https://doi.org/10.1016/j.neuroimage.2023.120253},
author = {Philipp Thölke and Yorguin-Jose Mantilla-Ramos and Hamza Abdelhedi and Charlotte Maschke and Arthur Dehgan and Yann Harel and Anirudha Kemtur and Loubna {Mekki Berrada} and Myriam Sahraoui and Tammy Young and Antoine {Bellemare Pépin} and Clara {El Khantour} and Mathieu Landry and Annalisa Pascarella and Vanessa Hadid and Etienne Combrisson and Jordan O’Byrne and Karim Jerbi},
}

@inproceedings{bajons2025mlsi,
  author    = {Bajons, R. and Kook, L.},
  title     = {Machine learning based statistical inference in sports analytics},
  booktitle = {Proceedings of the 39th International Workshop on Statistical Modelling (IWSM)},
  pages     = {218--222},
  year      = {2025},
}

@article{williamson20varimp,
    author = {Williamson, Brian D. and Gilbert, Peter B. and Carone, Marco and Simon, Noah},
    title = {Nonparametric Variable Importance Assessment Using Machine Learning Techniques},
    journal = {Biometrics},
    volume = {77},
    number = {1},
    pages = {9-22},
    year = {2020},
    month = {10},
    doi = {10.1111/biom.13392}
}

@Article{Strobl08varimp,
author={Strobl, Carolin
and Boulesteix, Anne-Laure
and Kneib, Thomas
and Augustin, Thomas
and Zeileis, Achim},
title={Conditional variable importance for random forests},
journal={BMC Bioinformatics},
year={2008},
month={Jul},
day={11},
volume={9},
number={1},
pages={307},
doi={10.1186/1471-2105-9-307}
}

@article{Shah20gcm,
author = {Rajen D. Shah and Jonas Peters},
title = {{The hardness of conditional independence testing and the generalised covariance measure}},
volume = {48},
journal = {The Annals of Statistics},
number = {3},
publisher = {Institute of Mathematical Statistics},
pages = {1514 -- 1538},
year = {2020},
doi = {10.1214/19-AOS1857},
}

@article{mews2022multistate,
  title={Multistate capture--recapture models for irregularly sampled data},
  author={Mews, Sina and Langrock, Roland and King, Ruth and Quick, Nicola},
  journal={The Annals of Applied Statistics},
  volume={16},
  number={2},
  pages={982--998},
  year={2022},
  publisher={Institute of Mathematical Statistics}
}

@article{koslik2025inference,
  title={Inference on the state process of periodically inhomogeneous hidden Markov models for animal behavior},
  author={Koslik, Jan-Ole and Feldmann, Carlina C and Mews, Sina and Michels, Rouven and Langrock, Roland},
  journal={The Annals of Applied Statistics},
  volume={19},
  number={4},
  pages={2724--2737},
  year={2025},
  publisher={Institute of Mathematical Statistics},
  note={\url{https://doi.org/10.1214/25-AOAS2107}}
}

@misc{bajons2025rGAX,
      title={Rethinking player evaluation in sports: Goals above expectation and beyond}, 
      author={Robert Bajons and Lucas Kook},
      year={2025},
      eprint={2509.20083},
      archivePrefix={arXiv},
      primaryClass={stat.AP},
      note={\url{https://arxiv.org/abs/2509.20083}} 
}

@article{chen2019package,
  title={Package ‘xgboost’},
  author={Chen, Tianqi and He, Tong and Benesty, Michael and Khotilovich, Vadim},
  journal={R version},
  volume={90},
  number={1-66},
  pages={40},
  year={2019},
  publisher={The R Foundation Vienna, Austria}
}

@article{wu2023evaluation,
  title={Evaluation of off-the-ball actions in soccer},
  author={Wu, Yifan and Swartz, Tim},
  journal={Statistica Applicata-Italian Journal of Applied Statistics},
  number={2},
volume = {35},
  year={2023}
}

@article{kook24comets,
    author = {Kook, Lucas and Lundborg, Anton Rask},
    title = {Algorithm-agnostic significance testing in supervised learning with multimodal data},
    journal = {Briefings in Bioinformatics},
    volume = {25},
    number = {6},
    year = {2024},
    doi = {10.1093/bib/bbae475},
}

@article{friedman2021package,
  title={Package ‘glmnet’},
  author={Friedman, Jerome and Hastie, Trevor and Tibshirani, Rob and Narasimhan, Balasubramanian and Tay, Kenneth and Simon, Noah and Qian, Junyang},
  journal={CRAN R Repositary},
  volume={595},
  pages={874},
  year={2021}
}

@article{Zou05elastic,
    author = {Zou, Hui and Hastie, Trevor},
    title = {Regularization and Variable Selection Via the Elastic Net},
    journal = {Journal of the Royal Statistical Society Series B: Statistical Methodology},
    volume = {67},
    number = {2},
    pages = {301-320},
    year = {2005},
    month = {03},
    doi = {10.1111/j.1467-9868.2005.00503.x}
}

@article{koslik2025tensor,
  title={Tensor-product interactions in {M}arkov-switching models},
  author={Koslik, Jan-Ole},
  journal={arXiv preprint arXiv:2507.01555},
  year={2025}
}

@inproceedings{chen2016xgboost,
  title        = {{XGBoost}: A scalable tree boosting system},
  author       = {Chen, Tianqi and Guestrin, Carlos},
  booktitle    = {Proceedings of the 22nd ACM SIGKDD International Conference on Knowledge Discovery and Data Mining},
  pages        = {785--794},
  year         = {2016},
  organization = {ACM},
  doi          = {10.1145/2939672.2939785}
}

@article{dempster2020rocket,
  title={ROCKET: exceptionally fast and accurate time series classification using random convolutional kernels},
  author={Dempster, Angus and Petitjean, Fran{\c{c}}ois and Webb, Geoffrey I},
  journal={Data Mining and Knowledge Discovery},
  volume={34},
  number={5},
  pages={1454--1495},
  year={2020},
  publisher={Springer}
}

@article{maruotti2009semiparametric,
  title={A semiparametric approach to hidden {M}arkov models under longitudinal observations},
  author={Maruotti, Antonello and Ryd{\'e}n, Tobias},
  journal={Statistics and Computing},
  volume={19},
  number={4},
  pages={381},
  year={2009},
  publisher={Springer}
}

@article{mcclintock2020uncovering,
  title={Uncovering ecological state dynamics with hidden {M}arkov models},
  author={McClintock, Brett T and Langrock, Roland and Gimenez, Olivier and Cam, Emmanuelle and Borchers, David L and Glennie, Richard and Patterson, Toby A},
  journal={{E}cology {L}etters},
  volume={23},
  number={12},
  pages={1878--1903},
  year={2020},
  publisher={Wiley Online Library}
}

@article{amoros2019continuous,
  title={A continuous-time hidden {M}arkov model for cancer surveillance using serum biomarkers with application to hepatocellular carcinoma},
  author={Amoros, Ruben and King, Ruth and Toyoda, Hidenori and Kumada, Takashi and Johnson, Philip J and Bird, Thomas G},
  journal={Metron},
  volume={77},
  number={2},
  pages={67--86},
  year={2019},
  publisher={Springer}
}

@article{zhang2019high,
  title={High-order hidden {M}arkov model for trend prediction in financial time series},
  author={Zhang, Mengqi and Jiang, Xin and Fang, Zehua and Zeng, Yue and Xu, Ke},
  journal={Physica {A}: {S}tatistical {M}echanics and its {A}pplications},
  volume={517},
  pages={1--12},
  year={2019},
  publisher={Elsevier}
}

@article{liu2012stock,
  title={Stock market volatility and equity returns: {E}vidence from a two-state {M}arkov-switching model with regressors},
  author={Liu, Xinyi and Margaritis, Dimitris and Wang, Peiming},
  journal={Journal of {E}mpirical {F}inance},
  volume={19},
  number={4},
  pages={483--496},
  year={2012},
  publisher={Elsevier}
}

@article{soper2020hidden,
  title={A hidden {M}arkov model for population-level cervical cancer screening data},
  author={Soper, Braden C and Nyg{\aa}rd, Mari and Abdulla, Ghaleb and Meng, Rui and Nyg{\aa}rd, Jan F},
  journal={{S}tatistics in {M}edicine},
  volume={39},
  number={25},
  pages={3569--3590},
  year={2020},
  publisher={Wiley Online Library}
}

%% or include bibliography directly:
% \begin{thebibliography}{}
% \bibitem[\protect\citeauthoryear{???}{???}]{b1}
% \end{thebibliography}
\newpage
%\begin{supplement}
%\appendix
\section*{Supplementary Material}
%\stitle{Supplementary Material}
%\sdescription{Supplementary Material to}
\setcounter{section}{0}
\setcounter{page}{1}
\setcounter{table}{0}
\setcounter{figure}{0}
\section{Initialization of the state distribution}

To obtain an informed initialization of the assignments between offensive and defensive players at the beginning of each play, we adopt a probabilistic matching approach based on the vertical distances between players. Specifically, for each defender \( i \) and each of the \( N = 5 \) potential offensive players \( j \), we compute a similarity score using a softmax over the negative distances,
\[
\delta_{ij} = \frac{\exp(-\alpha \, |\mathrm{y}^{(\text{def})}_i - \mathrm{y}^{(\text{att})}_j|)}{\sum_{k=1}^{n_{\text{att}}} \exp(-\alpha \, |\mathrm{y}^{(\text{def})}_i - \mathrm{y}^{(\text{att})}_k|)} \, ,
\]
where \( \mathrm{y}^{(\text{def})}_i \) and \( \mathrm{y}^{(\text{att})}_j \) denote the vertical positions of defender \( i \) and offensive player \( j \), respectively.  
The parameter \( \alpha > 0 \) controls the sharpness of the resulting distribution: higher values of \( \alpha \) yield nearly deterministic one-to-one assignments, while smaller values produce smoother, more uncertain matchings. In our implementation, we set \( \alpha = 1 \) based on heuristic considerations, balancing stability and interpretability of the initialization. This choice reflects the intuition that defenders tend to align with the nearest offensive player at the start of the play.  

\section{Lags}
\begin{figure}[b!]
    \centering
    \includegraphics[width=0.6\linewidth]{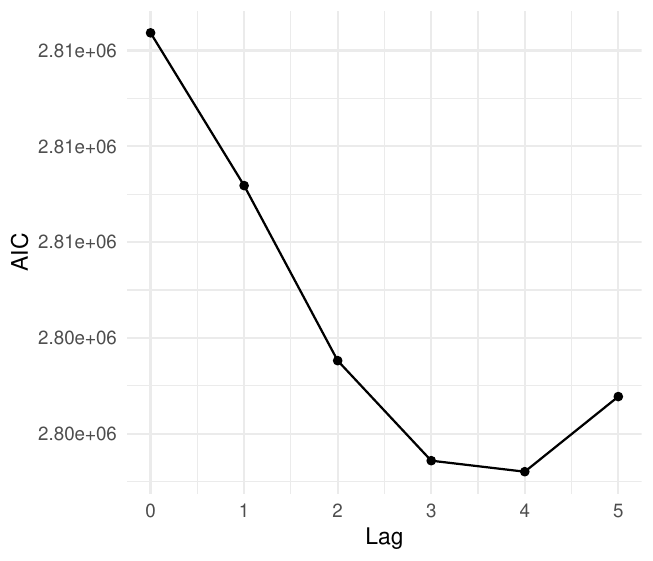}
    \caption{This figure displays the result of a preliminary analysis of a homogeneous HMM. On the x-axis, we displayed the lag $l$ (see Eq.~\ref{eq:lag}) and on the y-axis the corresponding AIC value.}
    \label{fig:lags}
\end{figure}

To determine an appropriate lag value $l$, we conducted a preliminary analysis using a homogeneous HMM. This decision was made separately from the full model design to avoid unnecessarily increasing the computational complexity. Lags ranging from $l=1, \ldots, 5$ were tested, and the corresponding models were evaluated based on the Akaike Information Criterion (AIC) (see Figure~\ref{fig:lags}). The model with the lowest AIC value was obtained for $l=4$ (similar to \citealp{wu2023evaluation} in soccer). This lag value was used in all subsequent analyses.

\section{Feature engineering and model features}
\label{app:features}

Prior to more involved feature engineering steps, we transform the coordinate system by redefining the x-variable as the x-distance to the end zone (such that all play directions are from right to left and the relevant end zone is at zero), and changing the direction, such that zero degrees represents heading straight towards the corresponding end zone. As mentioned in the main document, we use player features from five defensive and five offensive players. First, we standardized their $x$- and $y$-coordinates with respect to the football and ordered the players according to their $y$-coordinates, i.e., the first defender in our dataset is always the leftmost defensive player, while the first offensive player is always the rightmost one (offensive play direction is from left to right). A detailed list and description of all features used for modeling defensive coverage can be found in Table \ref{tab:variables}. %Note that initially, we extracted an even larger number of features containing information on the orientation and direction of the players, however due to small number data points, and resulting problems such as overfitting, we refrained from incorporating these features in the final analysis. 

\begin{table}[ht]
\centering
\tiny
\caption{Model features and descriptions.}
\label{tab:variables}
\resizebox{\linewidth}{!}{
\begin{tabular}{p{5cm} p{10cm}}
\hline
\textbf{Feature group} & \textbf{Description} \\
\hline

$\text{Convex hull area}^{\dagger}$ &
Area of the convex hull formed by player positions.
Includes:
\texttt{chull\_area\_o} (offense),
\texttt{chull\_area\_d} (defense). \\

$\text{Convex hull span (x-coordinate)}^{\dagger}$ &
Spread of the convex hull in x-coordinate direction.
Includes:
\texttt{chull\_x\_o} (offense),
\texttt{chull\_x\_d} (defense). \\

$\text{Convex hull span (y-coordinate)}^{\dagger}$ &
Spread of the convex hull in y-coordinate direction.
Includes:
\texttt{chull\_y\_o} (offense),
\texttt{chull\_y\_d} (defense). \\

$\text{Relative x-position of offensive players}^{\dagger}$ &
Relative x-coordinates of individual offensive players with respect to the football.
Includes:
\texttt{x\_rel\_player\_off1} to \texttt{x\_rel\_player\_off5}. \\

$\text{Relative x-position of defensive players}^{\dagger}$ &
Relative x-coordinates of individual defensive players with respect to the football.
Includes:
\texttt{x\_rel\_player\_def1} to \texttt{x\_rel\_player\_def5}. \\

$\text{Relative y-position of offensive players}^{\dagger}$ &
Relative y-coordinates of individual offensive players with respect to the football.
Includes:
\texttt{y\_rel\_player\_off1} to \texttt{y\_rel\_player\_off5}. \\

$\text{Relative y-position of defensive players}^{\dagger}$ &
Relative y-coordinates of individual defensive players with respect to the football.
Includes:
\texttt{y\_rel\_player\_def1} to \texttt{y\_rel\_player\_def5}. \\

$\text{Down}^{\dagger}$ &
Current down of the play (1st, 2nd, 3rd, or 4th down). \\

$\text{Quarter}^{\dagger}$ &
Game quarter in which the play occurs. \\

$\text{Yards to go}^{\dagger}$ &
Distance in yards required to achieve a first down. \\

$\text{Absolute yardline}^{\dagger}$ &
Absolute field position measured from a fixed reference (e.g., distance to end zone). \\

$\text{Pre-snap scores}^{\dagger}$ &
Scores of the home and visiting teams before the snap.
Includes:
\texttt{preSnapHomeScore},
\texttt{preSnapVisitorScore}. \\

$\text{Time remaining}^{\dagger}$ &
Number of seconds remaining in the current half. \\

\hline 

$\text{Maximum x distance}^{\dagger\dagger}$ &
Maximum x-distance covered by a player on the offensive or defensive team.
Includes:
\texttt{max\_x\_o},
\texttt{max\_x\_d}. \\

$\text{Maximum y distance}^{\dagger\dagger}$ &
Maximum y-distance covered by a player on the offensive or defensive team.
Includes:
\texttt{max\_y\_o},
\texttt{max\_y\_d}. \\

$\text{Total distance}^{\dagger\dagger}$ &
Total cumulative distance covered by the offensive and defensive team.
Includes:
\texttt{tot\_dist\_o},
\texttt{tot\_dist\_d}. \\

\hline 
$\text{Per play RE}^{\dagger\dagger\dagger}$ &
Predicted random effect of the play in the HMM model specification (see Section~\ref{sec:MC_dets}). \\

$\text{Sum of switches}^{\dagger\dagger\dagger}$ &
Sum of switches of latent guarding assignment as detected by the HMM. \\

$\text{\# of player changes}^{\dagger\dagger\dagger}$ &
Number of defensive players that at least switch one team the offensive guarding assignments during a play as detected by the HMM. \\

$\text{Average entropy}^{\dagger\dagger\dagger}$ &
Average entropy of the five defenders' decoded state distribution obtained from the HMM. \\

\hline
\hline
\multicolumn{2}{l}{$\dagger$ Pre motion features.} \\
\multicolumn{2}{l}{$\dagger\dagger$ Additional naive post motion features.} \\
\multicolumn{2}{l}{$\dagger\dagger\dagger$ Additional HMM features.}
\end{tabular}
}
\end{table}

\section{Random Effects Defense}
\begin{figure}[h!]
    \centering
    \includegraphics[width= \linewidth]{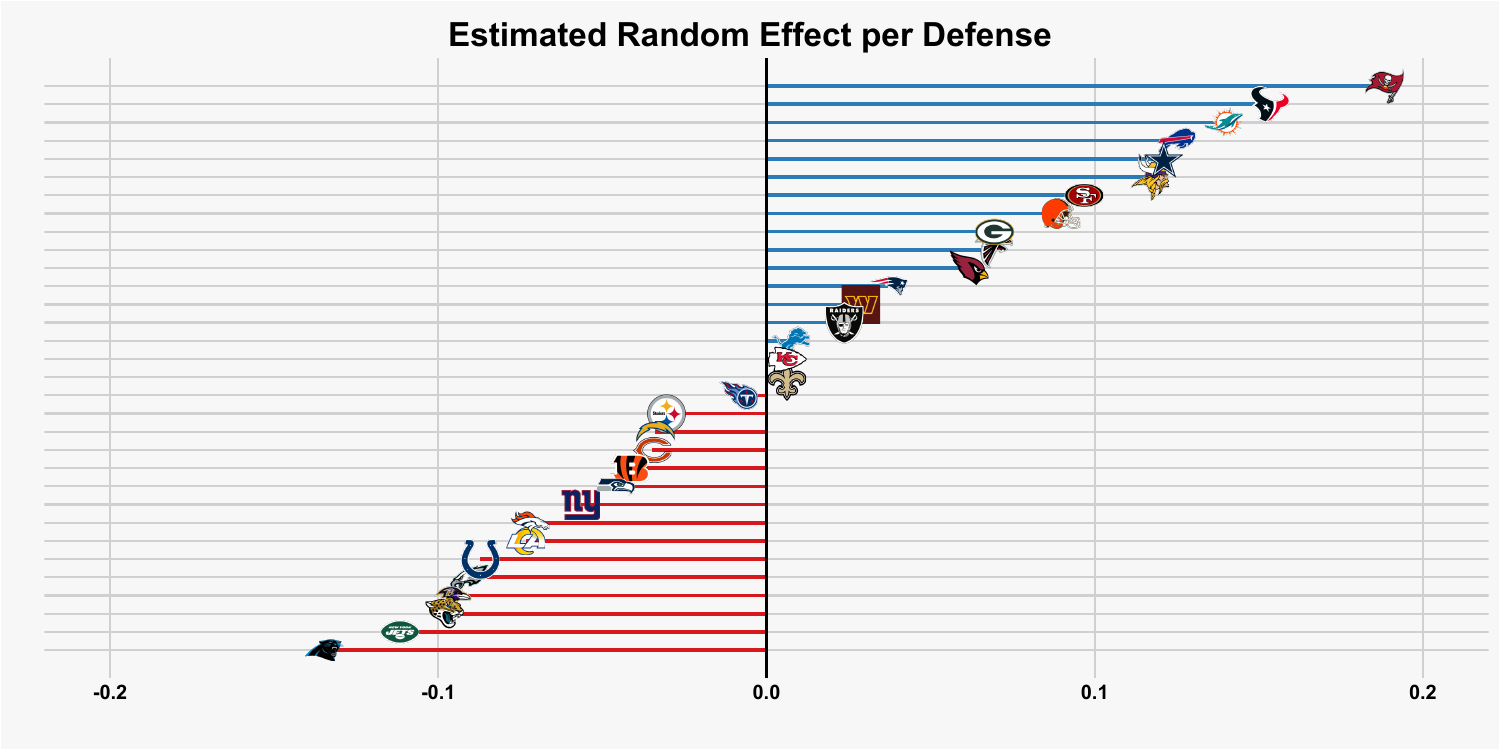}
    \caption{Predicted random effect for each defense. Smaller values correspond to a higher probability of staying in the respective state, i.e., continuing to covering a specific defender, and vice versa.}
    \label{fig:re_def}
\end{figure}

To account for systematic differences in defensive behavior across teams, random effects were incorporated into the transition probabilities of the HMM. This modeling approach enables team-specific deviations from the overall transition dynamics while maintaining a shared structure among individual defenders within each team, but also, consequently, each play. Figure~\ref{fig:re_def} presents the predicted random effects for all teams. Lower values indicate a higher probability of remaining in the current state (i.e., continuing to guard a specific offensive player), whereas higher values correspond to an increased likelihood of transitioning to another state.

%\end{supplement}

\end{spacing}
\end{document}